\documentclass[
aps,nofootinbib,
showpacs,showkeys,preprint
tightenlines,preprintnumbers,] {revtex4}

\usepackage{epsf,epsfig,subfigure,graphicx,amsmath,amssymb 
}
\usepackage{color}
\newcommand{\ie}{{\it i.e.~}}

\newcommand{\Qem}{Q_{\rm em}}

\newcommand{\eV}{\,\mathrm{eV}}
\newcommand{\Mp}{M_{\rm P}}

\newcommand{\Mg}{{M_{\rm GUT}}}

\newcommand{\Uoneem}{U(1)$_{\rm em}$}

\def\sw0{{$\sin^2\theta_W^0$}}

\def\fGUT{{\rm family-GUT}}

\newcommand{\Z}{{\bf Z}}

\def\E6{{\rm E_6}}

\def\EE8{{\rm E_8\times E_8'}}

\def\antiO{\Psi_{[A]}}
\def\antiD{\Psi^{[AB]}}

\def\nine{{\bf 9}}
\def\nineb{\overline{\bf 9}\,}

\def\tsixb{\overline{\bf 36}\,}
\def\tsix{{\bf 36}}

\def\one{\bf 1}

\def\five{\bf 5}
\def\ten{\bf 10}
\def\tenb{\overline{\bf 10}} 

\def\fiveb{\overline{\bf 5}}
\def\threeb{{\bf\overline{3}}}
\def\three{{\bf 3}}

\begin{document}

\draft

\title{\Large\bf Grand unfication models  from  SO(32) heterotic string }

\author{ Jihn E.  Kim}
\address
{Department of Physics, Kyung Hee University, 26 Gyungheedaero, Dongdaemun-Gu, Seoul 02447, Republic of Korea, and\\
Asia Pacific Center for Theoretical Physics (APCTP), Pohang 790-784, Korea}

\begin{abstract} 
Grand unification groups (GUTs) are constructed from SO(32) heterotic string via $\Z_{12-I}$ orbifold compactification. So far, most phenomenological studies from string compactification relied on $\EE8$ heterotic string, and this invites the SO(32) heterotic string very useful for future phenomenological
studies. Here, spontaneous symmetry breaking is achieved  by Higgsing of the anti-symmetric tensor representations of SU($N$). The anti-SU($N$) presented in this paper is a completely different class from the flipped-SU($N$)s from the spinor representations of SO($2N$). Here, we realize chiral representations: $\tsix\oplus 5\cdot\nineb $ for a SU(9) GUT and  $3\{{\ten}'_L\oplus {\fiveb}'_L\}$ for a SU(5)$'$ GUT. In particular, we confirm that the non-Abelian anomalies of SU(9) gauge group vanish and hence our compactification scheme achieves the key requirement. We also present the Yukawa couplings, in particular for the heaviest fermion, $t$, and lightest fermions, neutrinos. In the supersymmetric version, we present a scenario how supersymmetry can be broken dynamically via the confining gauge group SU(9). Three families in the visible sector are interpreted as the chiral spectra of SU(5)$'$ GUT.
 
\end{abstract}
\maketitle


\section{Introduction}\label{sec:Introduction}
\noindent

Grand unified theories (GUTs) attracted a great deal of attention ethetically because they provided unification of gauge couplings and charge quantization \cite{GQW74,PS73,GG74}. But there seems to be a fundamental reason leading to GUTs even at the standard model (SM) level.   With the electromagnetic and charged currents (CCs), the leptons need representations which are a doublet or bigger. A left-handed (L-handed) lepton doublet  $(\nu_e,e)$ alone is not free of gauge anomalies because the observed electromagnetic charges are not $\pm \frac12$. The anomalies from the fractional electromagnetic charges of the $u$ and $d$ quarks are needed to make the total anomaly to vanish \cite{Bouchiat72,Jackiw72}. In view of this necessity for jointly using both leptons and quarks to cancel gauge anomalies in the SM, we can view that GUTs are fundamentally needed in addition to the esthetic viewpoints.

In the SM, the largest number of parameters is from the Yukawa couplings which form the bases of the family structure. Repetition of fermion families in 4-dimensional (4D) field theory or  family-unified GUT (\fGUT) was formulated by Georgi \cite{Georgi79}, requiring un-repeated chiral representations while not allowing gauge anomalies. Some interesting \fGUT\,models are the spinor representation of SO(14) \cite{Kim80PRL,Kim81PRD} and ${\bf 84}\oplus 9\cdot\overline{\bf 9}$ of SU(9) \cite{Frampton80}.\footnote{For more attempts of \fGUT s, see references in \cite{FramptonKim20}.} While Refs. \cite{Kim80PRL,Kim81PRD,Frampton80}  do not provide interesting non-vanishing flavor quantum number, the SU(11) model \cite{Georgi79} allows a possibility for non-vanishing flavor quantum number such as U(1)$_{\mu-\tau}$ or U(1)$_{B-L}$ \cite{FramptonKim20}.

\bigskip
\noindent
On the other hand, the standard-like models from string have been the main focus of phenomenological activities for the ultraviolet completion of the SM  in the last several decades \cite{IKNQ, Munoz88, Lykken96, PokorskiW99, Cleaver99, Cleaver01, CleaverNPB, Donagi02,  Raby05, Donagi05, He05,  Donagi06, He06, Blumenhagen06, Cvetic06, Blumenhagen07, KimJH07, Faraggi07, Cleaver07, Munoz07,Lebedev08, Nilles08,Nibbelink07}. These models use the chiral specrum from the level--1 construction which leads to unification of gauge couplings \cite{Ginsparg87}.
So, the standard-like models from string compactification achieved the goal of gauge coupling unification and GUT theories from string  have not attracted much attention. Nevertheless, GUTs from strings \cite{Ellis89,KimKyae07} have been discussed sporadically  for anti-SU(5) \cite{DKN84} (or flipped SU(5) \cite{Barr82}), dynamical symmetry breaking \cite{Huh09,KimKyae19},  
and family unification \cite{Kimjhep15,Kim19Rp,PHFKim20}. In fact, \fGUT s  are much easier in discussing the family problem, in particular on the origin of the mixing between quarks/leptons, guiding to the progenitor mass matrix \cite{KimKim20} because the number of representations in \fGUT s is generally much smaller than in their (standard-like model) subgroups.

\bigskip
\noindent
In this paper we study \fGUT s from string compactification. So far, most  string compactification models used the $\EE8$ heterotic string in which a GUT with rank greater than 8 is impossible. In Ref. \cite{FramptonKim20}, to assign some non-vanishing $L_\mu-L_\tau$ family quantum number, only the \fGUT~SU(11) is chosen among the known \fGUT~models. The group SU(11) has rank 10   which cannot arise from compactification of $\EE8$. Therefore, firstly we fomulate the orbifold compactification \cite{DHVW1,DHVW2} of SO(32) heterotic string \cite{GHMR1} whose rank is 16.  
The SO(32) string compactification has been studied before \cite{Nibbelink07} but it did not include the GUTs. The GUT study is here for the first time.
  Then, we also attempt to accompany a hidden sector nonabelian group such that it provides a confining force toward breaking supersymmetry (SUSY) \cite{KimKyae19}.

\bigskip
\noindent
Among compactification schemes, we adopt the orbifold method. Among 13 possibilities listed in Ref. \cite{DHVW1}, we employ $\Z_{12-I}$ orbifold because it has the simplest twisted sectors. Twisted sectors are distinguished by Wilson lines \cite{INQ87}. The Wilson line in $\Z_{12-I}$ distinguishes three fixed points at a  twisted sector. Therefore, it suffices to consider only three cases at a twisted sector. In all the other orbifolds of Ref. \cite{DHVW1}, consideration of various possibilities of Wilson lines and the accompanying consistency conditions are much more involved. So, as the first step, in this paper we work with the  $\Z_{12-I}$ orbifold.  
 
\bigskip    
 \noindent 
 In Sec. \ref{sec:SU16}, we obtain the SU(16) subgroup of SO(32).   In Sec. \ref{section:Orb}, we recapitulate the orbifold methods used in this paper for an easy reference to  Sec. \ref{sec:SU9}.  
 Even though the computer program is put in Ref. 
 \cite{Vaudrevange11}, the GUT families are lacking from these programs. To our experience, there are not many possible  working GUTs and it is not possible to obtain them except from $\Z_{12}$. 
 In Sec. \ref{sec:SU9}, we list all possible massless SU(9) and SU(5)$'$ SUSY spectra. In Sec. \ref{sec:Singlets}, we list singlets which are needed to write down non-renormalizable couplings. In Sec. \ref{sec:Spontaneous}, we discuss symmetry breaking. Firstly, we comment on breaking the Georgi-Glashow SU(5)$'$ and discuss breaking SUSY by the SU(9) spectra.
Sec. \ref{sec:Conclusion} is a conclusion.  In Appendix, we list tables
possessing vector-like representations or no fields because of the cancelling-out phases,  from $T_3,T_4,T_1,T_2,$ and $T_5$ sectors.

\section{SU(16) subgroup}\label{sec:SU16}

\noindent
To discuss the family number in SU(5), the easiest way is to count the number of un-paired $\ten$'s.   The number of un-paired $\ten$'s  automatically determines  the number of un-paired $\fiveb$'s from the anomaly freedom. The anomaly unit  of $m$ completely anti-symmetric tensor representation  in SU($N$) is
\begin{eqnarray}
 &~{\cal A}([m])= \frac{(N-3)!(N-2m)}{(N-m-1)!(m-1)!}, \label{eq:anomaly}
 \end{eqnarray}
 where $[\overline{m}]=[N-m]$, \ie
\begin{eqnarray}
{\cal A}({[\bar{1}]})=-1,~{\cal A}\left({ [\bar{2}]}  \right)=-N+4,~ {\cal A}\left([\bar{3}] \right) = -\frac{(N-3)(N-6)}{2},\,{\rm etc.} \label{eq:anomaly1}
\end{eqnarray}
 Not to allow beyond $\three$ and $\threeb$ as quarks and anti-quarks of color SU(3),  we do not use higher dimensional representations for matter fields except the $m$ anti-symmetrized representations, $[{m}]$.
 
\bigskip 
\noindent
The adjoint representation {\bf 496} of SO(32) suggested in the heterotic string \cite{GHMR1} branches to the following SU(16) representations, 
\begin{equation}
\Phi^a_b \oplus \Phi^{[ab]} \oplus \Phi_{[ab]}, ~~(a, b=1,2,\cdots,16),~(n=16)\label{eq:AdjBranch}
\end{equation}
whose dimensions are $n^2={\bf 255}\oplus {\one}, \frac{n(n-1)}{2}={\bf 120}$, and $\frac{n(n-1)}{2}=\overline{\bf 120}$, respectively. In the orbifold compactification of SO(32), it will be easy to realize the representation $\Phi^{[ab]}$ and $\Phi_{[ab]}$ even at level 1 because they are anti-symmetric representations, and the key breaking pattern of \fGUT, \ie the separation of color SU(3)$_c$ and weak SU(2)$_W$, to the SM is possible by $\langle \Phi^{[45]}\rangle$ and $\langle\Phi_{[45]}\rangle$ of Eq. (\ref{eq:AdjBranch}). By restricting to the SU(16) subgroup of
SO(32), we exclude many possibilities of SO(32) where however we do not lose any chiral representation.
  
  \bigskip
  \noindent
Representations [1] and [2] have  the following matrix forms,
\begin{equation}
[1]\equiv \Phi^{[A]}=\begin{pmatrix}
\alpha_1\\[0.2em] \alpha_2\\[0.2em]  \alpha_3\\[0.2em]  \alpha_4\\[0.2em]  \alpha_5\\[0.2em]  f_6\\ \cdot \\[0.2em]  \cdot\\[0.2em]    \cdot\\[0.3em]  f_N\\[0.2em]
\end{pmatrix},\hskip 0.5cm
[2]\equiv \Phi^{[AB]}=\begin{pmatrix}
0, &\alpha_{12}, &\cdots,&\alpha_{15} & {\Big|}&\epsilon_{16},&\cdots,&\epsilon_{1N}\\[0.2em]
-\alpha_{12},& 0,&\cdots,&\alpha_{25}&{\Big|}&\epsilon_{26},&\cdots,&\epsilon_{2N}\\ 
\cdot&  \cdot&  \cdot&  \cdot&{\Big|}&  \cdot& \cdot& \cdot \\ 
\cdot&  \cdot&  \cdot& \alpha_{45}&{\Big|}&  \cdot& \cdot& \cdot \\ 
-\alpha_{15}, &-\alpha_{25},&  \cdots,&0&{\Big|}&\epsilon_{56},&\cdots,&\epsilon_{5N}\\[0.2em]
\hline
-\epsilon_{16},& -\epsilon_{26}\,&\cdots,&-\epsilon_{56}&{\Big|}&0,&\cdots,&\beta_{6N}\\ 
\cdot&  \cdot&  \cdot&  \cdot&{\Big|}&  \cdot& \cdot& \cdot \\ 
-\epsilon_{1N},& -\epsilon_{2N}\,&\cdots,&-\epsilon_{5N} &{\Big|}&-\beta_{6N},&\cdots,&0
 \end{pmatrix}  
 \end{equation}
 where  
[1] contains one $\five$, and [2] contains one $\ten$ of SU(5). The number of the SU(5)$_{\rm GG}$ families, \ie that of $\ten$ plus $\fiveb$, is counted by the number of $\ten$ minus the number of $\tenb$. 
The anomaly-freedom condition chooses the matching number of $\fiveb$'s. The numbers $n_1$ and $n_2$ for the vectorlike pairs $n_1{(\five\oplus\fiveb)}+ n_2{(\ten\oplus\tenb)}$ are not constrained by the anomaly freedom. Thus, we count the number of families  just by the net number of two index fermion representations in the SU(5)$_{\rm GG}$ subgroup. Because we allow only the
SM fields, the fundamental representations $\Phi^{[A]}$ and  $\Phi_{[A]}$, at the locations $f_6,\cdots, f_N$ of   Eq. (4), are also used to reduce the rank further by these VEVs.

\bigskip
\noindent
For the fundamantal representation in SU(9), we choose [1] as
\begin{equation}
\nine=(\underline{1\,0^8}).\label{eq:Vector9}
\end{equation}
In this case, obviously we have the following [2]
\begin{equation}
\tsix=(\underline{1\,1\,0^7}).\label{eq:Vector36}
\end{equation}
For spinors, however, it is more involved. An SO(18) spinor, which is $2^{8}(=256)$ dimensional, is chiral. In terms of SU(9) representations, let us define
\begin{eqnarray}
&&(\underline{+\,-^8}),\,(\underline{+++\,-^6}),\,(\underline{+^5\,----}),\,(\underline{+^7\,--}),\,(+^9),\\
{\rm SU(9):}&&\quad \nineb\quad\qquad~~\overline{\bf 84} \qquad\qquad~~ 
\overline{\bf 126}\qquad\qquad~~  \tsixb\qquad~\, \overline{\one} \label{eq:Spin9odd}
\end{eqnarray}
The complex conjugation of (\ref{eq:Spin9odd}) is
\begin{eqnarray}
&& (-^9), (\underline{++\,-^6}),\,(\underline{++++\,-^5}),\,(\underline{+^6\,---}),\,(\underline{+^8\,-}), \\
{\rm SU(9):}&&~~\one\qquad ~~{\bf 36} \quad\qquad\quad~{\bf 126}\qquad\qquad\quad {\bf 84}\qquad\quad~\, \nine \label{eq:Spin9even}
\end{eqnarray}
We defined the spinors in this way such that we include only even numbers of + signs inside
the SU(9) spinors. We will use the definitions given in Eqs. (5,6,7,8).

\bigskip
\noindent
We find that SU(9)  is the maximal subgroup of SU(16) from {\it string construction}, allowing  two indices anti-symmetric tensor fields. Its covering group is SO(18) which belongs to SO($4n+2$) groups allowing chiral spinors. 	The SO($4n+2$) groups were used for field theory GUTs \cite{SO10Georgi,SO10Fritzsch,Kim80PRL}. But, in string construction we cannot obtain spinors. Among branching of spinors to SU($2n+1$) representations,  we obtain at most two indices anti-symmetric tensor fields.

\section{Orbifold compactification}\label{section:Orb}

\noindent
Orbifolds are manifolds with identification of space points by discrete groups.\footnote{See, for example, a book presenting toolkits for orbifold compactification \cite{LNP954}.} This idea was used to reduce 6D internal space of 10D string to obtain 4D  light fields.  The internal 6D is so small that their details are shown up only through effective high dimensional interactions of the 4D light fields. Light fields appear in the untwisted sector $U$ and also in the twisted sector $T$
\cite{DHVW1,DHVW2}. Gauge groups are determined from $U$. In the untwisted sector $U$, spinor lattice points of $\EE8$ heterotic string satisfying $P^2=2$  arise also, but it is not so in the SO(32) heterotic string. In a sense, therefore in the SO(32) heterotic string than in the $\EE8$ heterotic string, it is easier to obtain gauge groups in the untwisted sector without the need for considering the spinor style lattice points.

\bigskip
\noindent
In the twisted sectors, there are fixed points and the fixed points can be distinguished by the Wilson lines which circle around the fixed points
\cite{INQ87}. 
In the most discussed $\Z_{6-II}$ and $\Z_{12-I}$ orbifolds, the number of fixed points are 12 and 3, respectively. Here, $\phi_s$ are given as $\phi_s=\frac16(3,2,1)$ and  $\phi_s=\frac{1}{12}(5,4,1)$, respectively, where each entry represents the two-dimensional torus of the internal six dimensions.
The cental number (in $\phi_s$) in $\Z_{6-II}$ and $\Z_{12-I}$ are 2 and 4, respectively, which mean that they have $\Z_{6/2}$ and $\Z_{12/4}$ symmetries, \ie both have the $\Z_3$ symmetry in the second torus.   Except in the second torus, we calculate the multiplicities by the direct product of the multiplicities in the remaining two tori in case there is no Wilson line, \ie the case $l=0$ of Table I in case of $\Z_{12-I}$.
 
  \bigskip
 \noindent
Toward the SU(9) \fGUT, we note that \cite{LNP954}
\begin{eqnarray}
\begin{array}{l}
\rm {\bf 1}.~ Matter~representations~ \Psi^{[ABCD]}~and~ \Psi_{[ABC]} ~ do~not~appear.\\ [0.2em]
\rm  {\bf 2}. ~Matter~\Psi_{[AB]}~and~\Psi_{[A]}~can ~appear~in~ the~untwisted\,(viz.~Eq.~(\ref{eq:AdjBranch}))~and~twisted~ sectors.  
\label{eq:conditions}\\ [0.2em]
\rm {\bf 3}.~ Among~mod~ integers,~choose ~only~one~ integer. 
\end{array}\label{eq:conditions}
\end{eqnarray}
In string compactifications, therefore, the number of families is counted by the number of the antisymmetric representation [2].
Matter in the untwisted sector $U_i$ occurs with $P\cdot V=\frac{N_i}{N}$.  For example,  $N_i=\frac{1}{12}(5,4,1)$ for $\phi_s$ of $\Z_{12-I}$ is shown in the second column of Table I.

\begin{table}[!ht]
\begin{center}
\begin{tabular}{|c|cccccccccccc|}
 \hline 
 &   & & & & &   $l$& $=$& & & &  &\\[-0.3em]
 $k$ &0&1&2 &3&4&5&6&7&8&9&10& 11\\ 
 \hline
 1&  3&3&3&3 &3&3&3 &3&3&3 &3&3\\[-0.2em] 
 2 &  3&3&3&3 &3&3&3 &3&3&3 &3&3\\[-0.2em] 
3 & 4& 1& 1& 4& 1& 1&  4& 1& 1& 4& 1& 1 \\ [-0.2em]
4& 9&1&1&1 &9&1&1 &1&9&1 &1&1 \\[-0.2em] 
   5 &  3&3&3&3 &3&3&3 &3&3&3 &3&3\\[-0.2em] 
6 & 16&1&1&4 &1&1&16 &1&1&4 &1&1\\
\hline
\end{tabular}
 \caption{ $\tilde{\chi}(k,l)$ in the $\Z_{12-I}$ orbifold. In the 4th row, we have 9\,1\,1\,1\,9\,1\,1\,1\,9\,1\,1\,1 instead of 27~9\, 9\, 9\, 27~9\, 9\, 9\,  27~9\, 9\, 9 of Ref. \cite{LNP696}. It is corrected in \cite{LNP954}.}\label{tab:Chis12I}
\end{center} 
\end{table}

 \bigskip
 \noindent
In the $k$-th twisted sector of $\Z_{N}$ orbifold, multiplicities  ${\cal P}_k$  is\footnote{$\tilde\chi(\theta^k,\theta^l)$ are presented in Ref. \cite{LNP954}.}
\begin{eqnarray}
{\cal P}_k   =\frac{1}{N}\sum_{l=0}^{N}
\tilde\chi(k,l)e^{i\,2\pi l\Theta_0},\label{eq:Multiplicity}
\end{eqnarray}
where $\tilde{\chi}(k,l)$ in the $\Z_{12-I}$ orbifold are listed in Table I and the phase angle $\Theta_0$ will be defined later.
The chirality is given by the first entry $s_0$ in $s$ with the even number of total `$-$'s in Eq. (\ref{eq:sDef}),  
\begin{eqnarray}
s=(s_0;\tilde{s})=(\ominus ~{\rm or}~\oplus\,;\pm,\pm,\pm),\label{eq:sDef}
\end{eqnarray}
where $s_0$ corresponds to L- or R- movers.
  
\begin{table}[!ht]
\begin{center} 
 \begin{tabular}{|c |ccccc|}
\hline
  &&&{\rm Multiplicity}&\\[-1.4em]
  &&&&&\\
  i  & ${\cal P}_k (0)$ &~ ${\cal P}_k (\pm\frac{\pi}{3})$& ${\cal P}_k ( \pm\frac{2\pi}{3} )$& ${\cal P}_k (\pm \frac{ \pi}{2} )$ &~~ ${\cal P}_k (\pi)$   \\[0.3em]
 \hline
 1   &3 &0&0&0&0\\
 2  & 3 &0 &0&0&0 \\
 3  &2 &0&1&0&0\\
 4  &3 &0&0&2&2\\
 5  &3&0&0&0&0\\
 6  &{4}  &2  &{3} &0 &{2} \\
\hline
  \end{tabular}
\end{center}
\caption{Multiplicities in the $k$-th twisted sectors of   $\Z_{12-I}$.  ${\cal P}_k({\rm angle})$ is calculaed with angle\,$=\frac{2\pi}{12}\cdot l$ in Eq. (\ref{eq:Multiplicity}).}\end{table}
In Table  II, multiplicities in the  $\Z_{12-I}$ orbifold are presented \cite{LNP954}.   Here, note that in $T_4$ we use $(9~1~ 1~ 1)^3$  instead of $(27~ 3~ 3~ 3)^3$. $T_5$ has the opposite chirality from those of $T_1$ and $T_2$, and we consider $T_7$ instead of $T_5$. In addition, note that $T_1, T_2$, and $T_5$ have nonvanishing multiplicities for ``angle'' = 0, and hence it is enough considering chiral spectra,  applying Eq. (\ref{eq:Multiplicity}) only to  $T_1, T_2$, and $T_7$. On the other hand, in $T_3, T_9, T_4, T_8,$ and $T_6$ more than one angle contribute to the chiral spectra, and accordingly  Eq. (\ref{eq:Multiplicity})  must be applied to all these sectors. The program of Ref. \cite{Vaudrevange11} is written in this way.

It is
proved  in this paper by explicitly calculating the number of chiral spectra.
In the twisted sector, the masslessness conditions are satisfied for the phases contributed by the left- and right-movers \cite{KimKyae07},
\begin{eqnarray}
2N_L^j\hat{\phi}_j +(P+kV)\cdot V -\frac{k}{2}V^2 =2\tilde{c}_k, ~{\rm L~movers},\label{eq:oscillatorL}
\end{eqnarray}
\begin{eqnarray}
2N_R^j\hat{\phi}_j - \tilde{s}\cdot\phi_s +\frac{k}{2}\phi_s^2  =2 {c}_k, ~{\rm R~movers},\label{eq:oscillatorR}
\end{eqnarray}
where $j$ denotes the coordinate of the 6-dimensional compactified space running over $\{1,\bar{1}\},\{2,\bar{2}\},\{3,\bar{3}\}$, and $\hat{\phi}^j=\phi_{s}^j\cdot {\rm sign}(\tilde\phi^j)$ with  ${\rm sign}({\phi}^{\bar j})=-{\rm sign}(\tilde\phi^j) $. 
The phase $\Theta_0$ in Eq. (\ref{eq:Multiplicity}) is   
\begin{eqnarray}
&\Theta_k  = \sum_j (N^j_L-N^j_R)\hat{\phi}^j -\frac{k}{2}(V_a^2-\phi_s^2)+(P+kV_a)\cdot V_a-(\tilde s +k\phi_s)\cdot\phi_s +{\rm integer}, \nonumber\\
&=-\tilde{s}\cdot\phi_s+\Delta_k,\label{eq:Phase}
\end{eqnarray}
where  $\Delta_k$ is
\begin{eqnarray}
&&\Delta_k = (P+kV_a)\cdot V_a-\frac{k}{2}(V_a^2-\phi_s^2)+\sum_j (N^j_L-N^j_R)\hat{\phi}^j\\
&&\qquad\equiv\Delta_k^0+\Delta_k^N.\label{eq:Deltak}
\end{eqnarray}
$V_a$ is the shift vector $V$ distinguished by Wilson lines $a,\,V_{0,+,-}$, and
\begin{eqnarray}
&&\Delta_k^0 =P\cdot V_a+\frac{k}{2}(-V_a^2+\phi_s^2) , \\
&&\Delta_k^N =\sum_j (N^j_L-N^j_R)\hat{\phi}^j.\label{eq:PhaseMore}
\end{eqnarray}
We choose $0<\hat{\phi}^j\le 1$ mod integer and  oscillator contributions due to $(N_L-N_R)$ to the phase can be positive or negative  with non-negative number $N_{L,R}\ge 0$. But each contribution to the vacuum energy, $N_{L}^j\hat{\phi}^j$ or  $N_{R}^j\hat{\phi}^j$, is nonnegative. One oscillation contributes one number in $\phi_s$.  With the oscillator $\hat{\phi}^j$, the vacuum energy is shifted to
\begin{eqnarray}
&(P+kV_a)^2+2\sum_j N_L^j \hat{\phi}^j = 2\tilde{c}_k\\
&(p_{\rm vec}+k\phi_s)^2+2\sum_j N_R^j \hat{\phi}^j = 2 {c}_k ,\label{eq:vacuumE}
\end{eqnarray}
where $2\tilde{c}_k$ and $ 2 {c}_k $ in the most discussed $\Z_{6-II}$ and $\Z_{12-I}$ orbifolds are
\begin{eqnarray}
\Z_{6-II}: &&\left\{\begin{array}{ll}
2\tilde{c}_k:  &
 ~ \frac{50}{36}(k=1),~ \frac{56}{36}(k=2),~ \frac{54}{36}(k=3),\\[0.3em]
 2 {c}_k: &
  ~ \frac{14}{36}(k=1),~ \frac{20}{36}(k=2),~ \frac{18}{36}(k=3),  
 \end{array}\right.\label{eq:Twist62}\\[0.4em] 
\Z_{12-I}: &&\left\{\begin{array}{ll}
2\tilde{c}_k:  &
 ~ \frac{210}{144}(k=1),~ \frac{216}{144}(k=2),~ \frac{234}{144}(k=3),~ \frac{192}{144}(k=4),~ \frac{210}{144}(k=5),~ \frac{216}{144}(k=6),\\[0.3em]
 2 {c}_k: &
  ~ \frac{11}{24}(k=1),~ \frac{1}{2}(k=2),~ \frac{5}{8}(k=3), ~\frac{1}{3}(k=4),~ \frac{11}{24}(k=5),~ \frac{1}{2}(k=6).  
 \end{array}\right.\label{eq:Twist121}
\end{eqnarray}
Note that $2\tilde{c}_k-2 {c}_k=1$ which is the required condition for ${\cal N}=1$ supersymmetry in 4D.  
 
\bigskip
 \noindent
The Wilson loop integral is basically the Bohm-Aharanov effect in the internal space of two-torus.  The complication arises at the points with $3a_3=0$ mod. integer, \ie at $T_{3,6}$ \cite{KimKyae07},\footnote{$T_9$ contains the CTP conjugate states of $T_3$.}  
 where the Bohm-Aharanov phase has to be taken into account explicitly. At  $T_{3,6}$ and also $U$,   for the (internal space) gauge symmetry we must require explicitly 
\begin{eqnarray}
(P+kV_0)\cdot a_3=0.\label{eq:WilsonCond}
 \end{eqnarray}
 We distinguish $T_3$ by $0,+$ and $-$ because the phase  $\Delta_k^0$ of Eq. (\ref{eq:PhaseMore}) contains an extra $\frac{k}{2}$ factor, but at $T_6$ the factor  $\frac{k}{2}$ is an integer. At $T_{1,2,4,5}$, we distinguish three fixed points just by $V_{0,+,-}$.
 
 \bigskip
 \noindent
 There is one point to be noted for $\Z_{12}$. If $N=$even, the $k=1,\cdots,\frac{N}{2}-1$ sectors provide the opposite chiralities in the  $k=N-1,\cdots,\frac{N}{2}+1$ sectors, 
 \begin{equation}
 T_k\leftrightarrow T_{N-k}.\nonumber
\end{equation}
Then, the corresponding phases of Eq. (\ref{eq:Phase}) compare as
 \begin{equation}
 e^{2\pi i \Theta_k}\leftrightarrow  e^{2\pi i \Theta_{N-k}}\label{eq:PhaseDiff}
\end{equation}
whose difference is $e^{2\pi i(N-2k)/12}$. Thus, if $2k=N$ then $T_{N-k}$ do not provide the charge conjugated fields of $T_k$, but they are identical. For $T_6$, therefore, we must provide the additional charge conjugated fields with an extra phase $e^{2\pi i(-10/12)}=e^{2\pi i(2/12)}$, the difference of $\hat{s}\cdot \phi_s$ for $\hat{s}=(+++)$ of R and (--\,--\,--) of L.  $T_3$ and $T_9$ form a doublet representation of $\Z_4$ and give the identical spectra. Here again, we must provide the additional charge conjugated fields. In $T_6$ and $T_9$ in our case,  we provide the charged conjugated fields by providing the extra phase $e^{2\pi i(2/12)}$.    
 
\subsection{Vacuum energy and multiplicity in the twisted sectors} 

\noindent 
 In the compactification of the $\EE8$ heterotic string,  spinors for rank 8 can contribute. But in the compactification of the SO(32) heterotic string, spinors in $U$ are not useful because $P=(\pm,\pm,\cdots,\pm)$ with sixteen entries gives $P^2=4$ instead of $P^2=2$. Only vector types are useful.  In the twisted sectors of $\Z_{12-I}$ orbifold, Wilson lines distinguish three fixed points in the second torus. At the $T_k$ twisted sector,   the three cases are
\begin{eqnarray}
T_k^{0,+,-}: ~kV_a=\left\{\begin{array}{l}kV\equiv kV_0\\[0.2em]
k(V+a_3)\equiv kV_+\\[0.2em]
k(V-a_3)\equiv kV_-.\end{array}\right.
\end{eqnarray}
Because $3a_3=0$ mod. integer, in the sectors with $k = \{3, 6, 9\}$, 0, +, and -- are not distingushed by the Wilson lines. But, Eq. (16) contains the factor $\frac12$ and hence $k = \{3, 9\}$
are distinguished by Wilson lines and $k = 6$ is not distinguished by Wilson lines. 

\bigskip
\noindent
We select only the even lattices shifted from the untwisted lattices, therefore, we consider even numbers for the sum of entries of each elements of $P$.
 
\bigskip 
\noindent
In the $k$-th twisted sector, the masslessness condition to raise the tachyonic vacuum energy to zero is  
\begin{eqnarray}
\left[(P+kV_a)^2 + 2\sum_j N_L^j \hat{\phi}^j \right]-2\tilde{c}_k=0,\label{eq:MasslessConV}
\end{eqnarray}
\begin{eqnarray}
\left[({\it p}+k\phi_s)^2 + 2\sum_j N_R^j \hat{\phi}^j \right]-2{c}_k=0 ,\label{eq:MasslessConS}
\end{eqnarray}
where $2\tilde{c}_k$ and $2 {c}_k$ are given in Eq.   (\ref{eq:Twist121}), and the brackets must be taken into account when oscillators contribute.
When the conditions (17) are satisfied, we obtain the SUSY spectra for which the chirality and multiplicity are calculated from $\Theta_0$ in the $k$-th twisted sector, from Eqs. (14) and (28)

\begin{eqnarray}
&\Theta_0   =-\tilde{s}\cdot\phi_s+k\, P\cdot V_0+\Delta_k^0+\Delta_k^N-\left( k\,p_{\rm vec}\cdot \phi_{s}+2\delta_k\right) , 
\end{eqnarray}
where $p_{\rm vec}, p_{\rm orb}$ and $\delta_k^{N}$ are given in Table \ref{tab:deltak}. $p_{\rm orb}^2$ saturates the   2nd line in Eq. (\ref{eq:Twist121}). $p_{\rm vec}$ in the right-moving sector mimics the lattice points $P$ in the left-moving sector.
 \begin{table} [!ht]
\begin{center}
\begin{tabular}{|c|c|ccc|cc| }
 \hline  
 Orbifold&\,Twisted Sector\,&  ~$k\,\hat{\phi}$~ & ~ $p_{\rm vec}$ & $p_{\rm orb}$& $-k\,p_{\rm vec}^{k\,\rm th}\cdot {\phi}_s$ & ~$\delta_k$~ \\[0.2em] \hline  
&$T_1$   & ~$(\frac{5}{12},\frac{4}{12} ,\frac{1}{12})$ &$(-1,0 ,0) $&$(\frac{-7}{12},\frac{+4}{12} ,\frac{+1}{12}) $ & $ \frac{5}{12}$ & $ \frac{1}{12}$   \\[0.3em]  
&$ T_2$    &~$(\frac{5}{6},\frac{4}{6}, \frac{1}{6})$&$(-1,0 ,0)$&   $(\frac{-1}{6},\frac{+4}{6} ,\frac{+1}{6})$ & $ \frac{10}{12}$& 0 \\[0.3em]   
\,$\Z_{12-I}$\, &~$ T_3$ ~ & ~$ (\frac{5}{4},\frac{4}{4} ,\frac{1}{4})$ &$ (-1,-1 ,-1)$
 & $(\frac{1}{4},\frac{0}{4} ,\frac{-3}{4})$& $ \frac{6}{12}$ &$\frac{3}{12}$ \\[0.3em]  
&$ T_4$   & ~$(\frac{5}{3},\frac{4}{3} ,\frac{1}{3})$&$ (-2,-1,0)$ &$(\frac{-1}{3},\frac{1}{3} ,\frac{1}{3})$& $ \frac{8}{12}$ &0  \\[0.3em]
 &$ T_5$ &~$(\frac{25}{12},\frac{20}{12} ,\frac{5}{12})$ &$(-2,-2,-1)$ &$(\frac{1}{12},\frac{-4}{12} ,\frac{-7}{12})$
  & $\times$   &$\times$ \\[0.3em]
&$ T_6$   &~$(\frac{5}{2},\frac{4}{2} ,\frac{1}{2}) $ &$(-2,-2,0) $ &$ (\frac{1}{2},\frac{0}{2} ,\frac{1}{2}) $ &0&0 \\[0.3em] 
 &$ T_7$   &~$(\frac{35}{12},\frac{28}{12} ,\frac{7}{12}) $ &$(-3,-1,0) $ &$ (\frac{-1}{12},\frac{4}{12} ,\frac{7}{12}) $   &$\frac{1}{12}$  &$\frac{1}{12}$ 
  \\[0.3em] 
&$ T_8$   &~$(\frac{40}{12},\frac{32}{12} ,\frac{8}{12}) $ &$(-3,-3,-1) $ &$ ( \frac{1}{3},\frac{-1}{3} ,\frac{-1}{3}) $ &$\frac{8}{12}$     &0   \\[0.3em] 
&$ T_9$   &~$(\frac{45}{12},\frac{36}{12} ,\frac{9}{12}) $ &$(-4,-3,-1) $ &$ (\frac{-3}{12},\frac{0}{12} ,\frac{-3}{12}) $   &$\frac{-3}{12}$    &$\frac{3}{12}$   \\[0.3em] 
 \hline
\end{tabular}
\end{center}
\caption{$H$ momenta, $p_{\rm orb}$, in the twisted sectors of $\Z_{12-I}$, Table 10.1 of \cite{LNP954}.   Requiring $(p_{\rm vec}+k\hat{\phi})^2=$(2nd line in Eq. (\ref{eq:Twist121})),  we have $p_{\rm vec}$ in the 4th column. In the last column, $\delta_k$ is shown, from which we have  the energy contribution from right movers $2\delta_k\ge 0$. } \label{tab:deltak}
\end{table}
 For $T_{1,2,5}$, multiplicities are determined by one angle as noted from Table I. Since the chiralities of $T_5$ fields have the opposite (or extra minus sign) to those of $T_1$ and $T_2$ fields, we consider $T_7$ (instead of $T_5$) to have the same chirality as those of $T_1$ and $T_2$. For $T_n\,(n=3,4,6)$, more than one angle contribute to the spectra, and hence we consider $T_{8}$ and $T_9$ also. In Table III, we also listed $\delta_k$ for which only $T_{1,3,5,9}$ have nonzero contributions. For $T_9$, the extra vacuum energy is needed. For the others saturating the 2nd line of Eq. (\ref{eq:Twist121}), we consider $\delta_k^N=2\delta_k$ if $p_{\rm orb}^2-(k\hat{\phi}^2)^2>0$ since we considered too much shifts in  $p_{\rm orb}$'s.

\begin{eqnarray}
\Delta_k^0 = \frac{k}{2}(\phi_s^2-V_a^2) , \\
\Delta_k^N =2\sum_j  N^j_L\,\hat{\phi}^j , \label{eq:DeltaNkA}\\
\delta_k^N =2\sum_j   N^j_R\,\hat{\phi}^j.\label{eq:del}
\end{eqnarray}
As an example, consider the $T_3$ sector.  Note that $(p_{\rm vec}+3\phi_s)^2=(\frac{1}{4},0,\frac{-3}{4})^2=\frac58$ with $p_{\rm vec}=(-1,-1,-1)$, which saturates $2c_3 = \frac5
8$ of Eq. (19).  Hence,  the  $N_R$ contribution is $0$. If we choose   $0<\hat{\phi}^j\le 1$ mod integer, not using $\hat{\phi}^{\overline{j}}$, oscillator contributions due to $(N_L-N_R)$ can be in principle positive or negative.  We used $p_{\rm vec}\cdot \phi_s=\frac{-10}{12}$
 as shown in Table IV because $p_{\rm vec}$ is already listed in the $k^{\rm th}$ twisted sector. 

\bigskip
\noindent
We will select only the even lattices shifted from the untwisted lattices. They form even numbers if the entries of each elements of $P$ are added.  
 In the tables, we list SU(9)   and SU(3)$'$ non-singlets and columns are ordered according to
9-*-* \begin{eqnarray}
&\Theta_{\rm Group}  =-\tilde{s}\cdot\phi_s -k\,p_{\rm vec}^{ k\,\rm th}\cdot \phi_{s} + k\,P\cdot V_0+\frac{k}{2}(\phi_s^2-V_0^2)+\Delta_k^N-  \delta_k^N ,\label{eq:PhaseA}
\end{eqnarray}
where 
\begin{eqnarray}
\delta_k^N=2\delta_k .\label{eq:deltakN}
\end{eqnarray}

\section{Family unification with SU(9) GUT}
\label{sec:SU9}

\bigskip
\noindent
 We anticipated to achieve the anomaly-free key spectra needed for SU(9) \fGUT,
\begin{eqnarray}
3\antiD+12\antiO+  \cdots,\label{eq:TENinSU8}
\end{eqnarray}
where$\cdots$ contain vectorlike pairs and singlets. Since it is impossible to obtain high dimensional representations $\Psi^{[ABC]}$ and $\Psi^{ABCD}$ from orbifold compactification, the family number is counted by the number of $\Psi^{[AB]}\equiv {\bf 36}$. We are interested in obtaining three chiral families.  The chiral representations are represented by $\Psi$'s, and   vectorlike reresentations  are represented by $\Phi$ which contain candidates for the Higgs bosons.

\bigskip
\noindent
The orbifold conditions, toward a low energy 4D effective theory,  remove some weights of the original ten dimensional SU(16) weights. The remaining ones constitute the gauge multiplets and matter fields in the untwisted sector in the low energy 4D theory. Therefore, the weights in the untwisted sector  $U$  must satisfy $P^2=2$. Because the rank of U(16) is 16, spinors with $P^2=2$ are not available. Orbifold conditions produce singularities. They are typically represented in three two-dimensional tori. A loop of string can be twisted around these singularities and define twisted sectors $T_k^{0,+,-}\,(k=1,2,\cdots,11)$. Twisting can introduce additional phases. Since $T_{12-k}$ provides the anti-particles of $T_k$, we consider only $T_k$ for $k=1,2,\cdots,6$. $T_6$ contains both particles and anti-particles. $T_6$, not affected by Wilson lines, is like an untwisted sector. It contains the antiparticles also as in $U$.\footnote{Actually, $T_{12}$ can be viewed as $U$.}

\bigskip
\noindent 
The shift vector $V_0$ and Wilson line $a_3$ are restricted to satisfy the $\Z_{12-I}$ orbifold conditions, 
\begin{eqnarray} 
&12 (V_0^2-\phi_s^2)= 0~\textrm{mod even integer},\\
& 12(V_0\cdot a_3)=0~\textrm{mod even integer},\\ 
&12|a_3|^2 =0~\textrm{mod even integer}. \label{eq:Conditions}
\end{eqnarray}
Here, $a_3\,(=a_4)$ is chosen to allow and/or forbid some spectra,
and is composed of fractional numbers with the integer multiples of $\frac13$ because the second torus has the $\Z_3$ symmetry.  
Toward SU(9) non-singlet spectra  in the $\Z_{12-I}$ orbifold from SO(32) heterotic string,   we choose the following model,
\begin{eqnarray}
 \begin{array}{l}  V_0=\left(\frac{1}{12},\,\frac{1}{ 12},\,\frac{1}{12},\,\frac{1}{12},\,\frac{1}{12},\,\frac{1}{12},\,\frac{1}{ 12},\,\frac{1}{12},\,\frac{1}{12};\, \frac{3}{12},\,  \frac{6}{12}\,;\, \frac{6}{12},\,\frac{6}{12},\,\frac{6}{12},\,\frac{6}{12},\,\frac{6}{12} \right) ,~V^2_0=\frac{234}{144}\to \frac{-54}{144},    \\[0.4em]
 V_+=\left( \frac{+1}{12},\, \frac{+1}{12},\, \frac{+1}{12}, \, \frac{+1}{12},\, \frac{+1}{12}, \,\frac{+1}{12},\, \frac{+1}{12},\, \frac{+1}{12}, \,\frac{+1}{12};\, \frac{+3}{12},\, \frac{+2}{12}\,;\, \frac{+10}{12},\, \frac{+10}{12}\,;\,\frac{+10}{12},\, \frac{+10}{12},\, \frac{+10}{12}\right),~V_+^2 =\frac{522}{144}\to \frac{-54}{144},\\ [0.4em] 
 V_-=\left( \frac{+1}{12},\, \frac{+1}{12},\,\frac{+1}{12},\,\frac{+1}{12},\,\frac{+1}{12},\,\frac{+1}{12},\,\frac{+1}{12},\,  \frac{+1}{12}, \,\frac{+1}{12};\,  \frac{+3}{12},\, \frac{+10}{12}\,;\, \frac{+2}{12},\, \frac{+2}{12},\,\frac{+2}{12} ,\, \frac{+2}{12},\, \frac{+2}{12} \right),~V_-^2 =\frac{138}{144} %
   \end{array} \label{eq:Model}
\end{eqnarray}
where 
\begin{eqnarray}
a_3=a_4=\left(0^9; \,0,\,\frac{-1}{3}\,;\,\frac{1}{3},\,\frac{1}{3},\,\frac{1}{3},\,\frac{1}{3},\,\frac{1}{3}\right) . 
\end{eqnarray}
 The   R-hand weights are
\begin{eqnarray}
&\phi_s^2=\frac{42}{144}.
\end{eqnarray}
 Shifted lattices by Wilson lines are given by $V_+$ and $V_-$,

  \bigskip
  \noindent
The order of presentation is  $U,T_3,T_4,T_1,T_2,$ and $T_5$ which contain chiral spectra.  
Finally, we present $T_6$ which contains only vector-like pairs. $V_0$ is the most important shift vector of $\Z_{12-I}$.
 In this paper, we are interested in obtaining chiral spectra and hence do not discuss $T_6$ which gives only vector-like spectra.
 
 \bigskip
 \noindent 
 We use the following notations: $V_{0,+,-}$ represent (left-hand or gauge goroup) shift vectors and $P_{\rm Group}$ (or sometimes just $P$ if no confusion arises) is the lattice point  in the SU(16) group space.
 
\subsection{Untwisted sector $U$} 
\noindent
In $U$, we find the following nonvanishing roots of SU(9)$\times$SU(5)$'\times$U(1)$^4$, 
\begin{eqnarray}
 {\rm SU(9)~gauge~ multiplet:}~~& P\cdot V =0{\rm~mod.~ integer}
 \nonumber\\
{\rm SU(9)}:& \left\{\begin{array}{l} P=(\underline{+1\, -1\, 0\, 0\, 0\, 0\, 0\,  0\, 0\,};0\, 0\,;0\, 0\, 0\,  0\, 0)\end{array}\right.\\[0.3em]
 {\rm SU(5)'~gauge~ multiplet:}~~& P\cdot V =0{\rm~mod.~ integer ~and~}P\cdot a_3=0 {\rm~mod.~ integer}\nonumber\\
 {\rm SU(5)}': & \left\{\begin{array}{l}  
 P=(0^{9};\,0^2;\, \underline{1\, -1\,0\,0\,0}\,).
   \end{array} \right.  \label{eq:Blwa3}
\end{eqnarray}
 For tensor notations, we use $A$ for SU(9) representations and $\alpha$ for SU(5)$'$ representations. 
In addition, there exists U(1)$^4$ symmetry. 
The non-singlet matter fields are
\begin{eqnarray}
  {\rm SU(9)~and/or~SU(5)'~matter~ multiplet:}~&P\cdot V =\frac{1,~4,~5}{12},P\cdot a_3=0 {\rm~mod.~ integer}
  \label{eq:Untwisted}
\end{eqnarray}
  The conditions (\ref{eq:Untwisted}) allows the  $P^2=2$ lattice shown in Table IV. 
The four entry set $s$ is the $s^2=2$ right-hand spin lattice, $s=( \ominus~{\rm or}~\oplus;\hat{s})$ with every entry being interger multiples of $\frac12$. In the following + and -- represent $\frac{+1}{2}$ and  $\frac{-1}{2}$, respectively. Three entry set in the right-hand sector is also used
  \begin{eqnarray}
  \hat{\phi}_s=\left(\frac{5}{12},\frac{4}{12},\frac{1}{12}\right).
  \label{eq:phishat}
  \end{eqnarray}
  
 \bigskip 
 \begin{table} [!h]
\begin{center}
\begin{tabular}{|c|c|c|c| c| }
 \hline  
$U_i$ &\,$P$\,&\,Tensor~form\,&\,Chirality\,&\,$[p_{\rm spin}]~(p_{\rm spin}\cdot\phi_s)$\, \\[0.1em] \hline  
   ~$U_1\,(p\cdot V=\frac{5}{12})$~  &--&~None~ &-- &-- \\   
    $U_2\,(p\cdot V=\frac{4}{12})$ & $(\underline{1~0^7};1~0 ~0;0^5)$ &   ~$\Psi^{A*}$~&L&$[\ominus;++-]~~\left( \frac{+4}{12} \right)$  \\ 
    $U_3\,(p\cdot V=\frac{1}{12})$  & --&   ~None~&-- &--   \\[0.1em]  
\hline
\end{tabular}
\end{center}
\caption{There is a ${\color{red}\nineb_R( {\Psi}^{A*}_R)}$   in view of Eq. (\ref{eq:Vector9}) in the twisted sector convention as the 12th twisted sector.  Chirality is read from the circled sign in     $s=(\ominus~{\rm or}~\oplus;\pm,\pm,\pm)$ where $\pm$ represents $\pm\frac12$.    $s=(\ominus;++-)=(\ominus;p_{\rm spin})$ gives chirality L ($\ominus$) because $P\cdot V_0=p_{\rm spin}\cdot\hat{\phi}_s=\frac{4}{12}$ where $\hat{\phi}_s$  is shown in Eq. (\ref{eq:phishat}). The convention on the chirality in $U$ as the 12th twisted sector defined from $T_{1,2,\cdots,6}$ is the opposite of $\ominus$ or $\oplus$. In the same way, we take the opposite chirality from $\ominus$ or $\oplus$ in the twisted sector $T_5$, since the 1st entry in $5\hat{\phi}_s$ exceeds 2. }\label{tab:Unt}
\end{table}

\noindent
In the multiplicity calculation in $\Theta_k$ in the $k^{\rm th}$ twisted sector, there is a factor $\frac12$ between the lattice shifts by Wilson lines. This is taken into account by changing the chiralities in $U$ as the $N^{\rm th}$ twisted sector in Table \ref{tab:Unt}.

\subsection{Twisted sector $T_3 \,( \delta_3 =\frac{3}{12})$}
 \noindent
 Even though the Wilson lines cannot distinguish the fixed points, we consider $V_+$ and $V_-$ also as if Wilson lines distinguish fixed points. 
In $T_3$, $3V_{0,+,-}$ become
\begin{eqnarray}
3V_0&=&\left( (\frac{+3}{12})^{9}\,;\,\frac{+9}{12}\, \frac{+18}{12};\, (\frac{+18}{12})^5 \, \right),\label{eq:3Vzero}\\ 
3 V_+&=&\left( (\frac{+3}{12})^{9}\,;\, \frac{+9}{12}\, \frac{+6}{12}\,;\, (\frac{+30}{12})^5\, \right),\\ 
3V_-&=&\left( (\frac{+3}{12})^{9}\,;\, \frac{+9}{12}\, \frac{+30}{12}\,;\, (\frac{+6}{12})^5\, \right).
\label{eq:3Tminus}
\end{eqnarray}
 
 \begin{table}[!ht]
\begin{center}
\begin{tabular}{|cc|c|cc|ccc|cc|}
 \hline &&&&&&&&& \\[-1.15em]
  Chirality  &   $\tilde s$& $-\tilde{s}\cdot\phi_s$& $-k\,p_{\rm vec}^{ k\,\rm th}\cdot \phi_s$ &$k\,P_{9}\cdot V_0$& $(k/2)\phi_s^2$,&$  -(k/2) V_0^2 $,& ~~$\Delta_{3}^N,~-\delta_3^N
  $   &~~$\Theta_{3},$& Mult. of SU(9)  \\[0.15em]
 \hline &&&&&&& && \\[-1.15em]
$\ominus=L$& $(---)$  &  $\frac{+5}{12}$ & $~~\frac{+6}{12},$& $\frac{-6}{12}$ &$\frac{63}{144}  $  &$\frac{+81}{144}$ & $ \frac{0}{12},~~\frac{- 6}{12}$&$\frac{-1}{12}  $& 0  \\ [0.1em]
 \hline &&&&&&&&& \\[-1.25em]
 $\ominus=L$ & $  (-++)$&  $0$ &$~~\frac{+6}{12},$& $\frac{-6}{12}$  &$\frac{63}{144} $  &$\frac{+81}{144}$& $ \frac{0}{12},~~\frac{-6}{12}$&$\frac{-6}{12}  $& 0\\[0.15em]  
 \hline  &&&&&&&& & \\[-1.25em]
$\ominus=L$&  $ (+-+)$  & $\frac{-1}{12}$ &$~~\frac{+6}{12}$&$\frac{-6}{12}$  &$\frac{63}{144} $  &$\frac{+81}{144}$&$ \frac{0}{12},~~\frac{-6}{12}$& $\frac{-7}{12}  $&  0 \\[0.15em]  
\hline &&&&&&&&&  \\[-1.25em]
 $\ominus=L$&  $  (++-)$ &$\frac{-4}{12}$&$~~\frac{+6}{12}$& $\frac{-6}{12}$  &$\frac{63}{144} $  &$\frac{+81}{144}$& $ \frac{0}{12},~~\frac{-6}{12}$&$\frac{-10}{12}  $&  0 \\[0.1em]  
 \hline &&&&&&&&& \\[-1.25em]
$\oplus=R$& $(+++)$  &  $\frac{-5}{12}$ & $~~\frac{+6}{12}$ & 
$\frac{-6}{12}$  & $\frac{63}{144}$ & $\frac{+81}{144}$ & $ \frac{0}{12},~~\frac{-6}{12}$ & $\frac{+1}{12}$  &0  \\ [0.1em]
 \hline &&&&&&&&& \\[-1.25em]
 $\oplus=R$ & $ (+--)$&  $0$ & $~~\frac{+6}{12}$ & $\frac{-6}{12}$ & $\frac{63}{144} $  & $\frac{+81}{144}$ &$ \frac{0}{12},~~\frac{-6}{12}$& $\frac{+6}{12}  $& 0  \\[0.15em] 
  \hline  &&&&&&&&&  \\[-1.25em]
$\oplus=R$&  $ (-+-)$  & $\frac{+1}{12}$ &$~~\frac{+6}{12}$&$\frac{-6}{12}$ &$\frac{63}{144} $  &$\frac{+81}{144}$& $ \frac{0}{12},~~\frac{-6}{12}$&$\frac{+7}{12}  $& 0   \\[0.15em] 
 \hline &&&&&&&&&  \\[-1.25em]
 $\oplus=R$&  $  (--+)$ &$\frac{+4}{12}$&$~~\frac{+6}{12}$& $\frac{-6}{12}$ &$\frac{63}{144} $  &$\frac{+81}{144}$&$ \frac{0}{12},~~\frac{-6}{12}$&$\frac{+10}{12}  $&  0 \\[0.15em] 
\hline
\end{tabular}
\end{center}
\caption{Two indices  spinor-form from $T_3^0$:   All states are projected out. }  
\end{table}
    
 \begin{table}[!ht]
\begin{center}
\begin{tabular}{|cc|c|cc|ccc|cc|}
 \hline &&&&&&&&& \\[-1.15em]
  Chirality  &   $\tilde s$& $-\tilde{s}\cdot\phi_s$& $-k\,p_{\rm vec}^{ k\,\rm th}\cdot \phi_s$ &$k\,P_{9}\cdot V_0$& $(k/2)\phi_s^2$,&$  -(k/2) V_0^2 $,& ~~$\Delta_{3}^N,~-\delta_3^N
  $   &~~$\Theta_{3},$& Mult. of SU(9)  \\[0.15em]
 \hline &&&&&&& && \\[-1.15em]
$\ominus=L$& $(---)$  &  $\frac{+5}{12}$ & $~~\frac{+6}{12},$& $\frac{-6}{12}$ &$\frac{63}{144}  $  &$\frac{+81}{144}$ & $ \frac{0}{12},~~\frac{- 6}{12}$&$\frac{-1}{12}  $& 0  \\ [0.1em]
 \hline &&&&&&&&& \\[-1.25em]
  $\oplus=R$& $(+++)$  &  $\frac{-5}{12}$ & $~~\frac{+6}{12}$&$\frac{-6}{12}$  &$\frac{63}{144}$ &$\frac{+81}{144}$&$ \frac{0}{12},~~\frac{-6}{12}$& $\frac{+1}{12}$  &0   \\[0.15em] 
\hline
\end{tabular}
\end{center}
\caption{The abbreviated form of Table V.  } 
\end{table}
 
\noindent$\bullet$   Two indices spinor-form from $T_3^0$:
 the spinor forms satisfying the mass-shell condition $(P+3V_0)^2=\frac{234}{144}=\frac{13}{8}$  and the Wilson-line condition $12(P+3V_0)\cdot a_3= 0$ are possible for SU(9):
  \begin{equation}
 P_{9}=\left(\underline{++-^{7}};-,\frac{-3}{12};(\frac{-3}{12})^5\right),
 \end{equation}
 but all states are projected out as shown in Table V. Table V can be abbreviated by the top entries of L and R fields as shown in Table VI. If the multiplicity is zero as in Table V, below it is stated as ``projected out''.
 
 \bigskip
\noindent$\bullet$   One index spinor forms from $T_3^0$:
 the spinor forms satisfying  $(P+3V_0)^2=\frac{234}{144}=\frac{13}{8}$  and $12(P+3V_0)\cdot a_3= 0$ are possible for SU(9):
  \begin{equation}
 P_{9}=\left(\underline{+\,-^{8}};-,\frac{-3}{2};(\frac{-3}{2})^5\right),\label{eq:NinebAbar}
 \end{equation}
 for which the massless fields are projected out.  
    
\bigskip
\noindent$\bullet$
Two indices spinor form for $T_3^+$:  the spinor forms satisfying $(P+3V_+)^2=\frac{234}{144}=\frac{13}{8}$ and $12(P+3V_+)\cdot a_3=0$
are possible for SU(9):
\begin{equation}
P_9=\left(\underline{++-^7};--;(\frac{-5}{2})^5\right)
\end{equation}
for which massless fields are projected out.  

\bigskip
\noindent$\bullet$ One index spinor-form from $T_3^+$:
the spinor forms satisfying  $(P+3V_+)^2=\frac{234}{144}=\frac{13}{8}$ and $12(P+3V_+)\cdot a_3=0$
are possible for SU(9): 
\begin{equation}
P_9=\left(\underline{+-^8};\frac{-3}{2} -;(\frac{-5}{2})^5\right)
\end{equation}
for which massless fields are projected out.

\bigskip
\noindent$\bullet$   Two indices spinor-form from $T_3^-$:
 the spinor form 
  \begin{equation}
 P_{9}=\left(\underline{++-^{7}};-\,\frac{-5}{2};-^5\right).
 \end{equation}
 gives  $(P+3V_0)^2=\frac{234}{144}$. 
 The massless fields are projected out since $\Theta_3=$ in the top row is $\frac{-1}{12}$.  
  
\bigskip
\noindent$\bullet$   One index spinor-form from $T_3^-$:
 the spinor form 
  \begin{equation}
 P_{9}=\left(\underline{+\,-^{8}};\frac{-3}{2},-;-^5\right).
 \end{equation}
 gives  $(P+3V_0)^2=\frac{234}{144}$ and massless fields are projected out.

\subsection{Twisted sector $T_9 \,( \delta_3 =\frac{3}{12})$}
 \noindent
Since chiral fields in $T_3$ are obtained from two angles, we check $T_9$ also not to miss the possibility of different linear combinations, of angles from $T_3$ and $T_9$, contributing to the massless fields.   In $T_9$, we have
\begin{eqnarray}
9V_0&=&\left( (\frac{+9}{12})^{9}\,;\,\frac{+27}{12}\, \frac{+54}{12}\,;\, (\frac{+54}{12})^5 \, \right), \\ 
9 V_+&=&\left( (\frac{+9}{12})^{9}\,;\, \frac{+27}{12}\, \frac{+18}{12}\,;\, (\frac{+90}{12})^5\, \right), \\ 
9V_-&=&\left( (\frac{+9}{12})^{9}\,;\, \frac{+27}{12}\, \frac{+90}{12}\,;\, (\frac{+18}{12})^5\, \right).
\label{eq:3Tminus}
\end{eqnarray}
Note that we will provide an extra phase $e^{2\pi i(2/N)}$ as commented below Eq. (\ref{eq:PhaseDiff}).

\bigskip
\noindent$\bullet$   Two indices spinor-form from $T_9^0$:
 the spinor forms satisfying the mass-shell condition $(P+9V_0)^2=\frac{234}{144}=\frac{13}{8}$  and the Wilson-line condition $12(P+9V_0)\cdot a_3= 0$ are possible for SU(9):
  \begin{equation}
 P_{9}=\left(\underline{\frac{-3}{2}\,\frac{-3}{2}\,-^{7}};\frac{-5}{2},\frac{-9}{2};(\frac{-9}{2})^5 \right).
 \end{equation} 
The massless fields are shown in Table \ref{tab:AbbTnine}. 
 
 \begin{table}[!ht]
\begin{center}
\begin{tabular}{|cc|c|cc|ccc|cc|}
 \hline &&&&&&&&& \\[-1.15em]
  Chirality  &   $\tilde s$& $-\tilde{s}\cdot\phi_s$& $-k\,p_{\rm vec}^{ k\,\rm th}\cdot \phi_s$ &$k\,P_{9}\cdot V_0$& $(k/2)\phi_s^2$,&$  -(k/2) V_0^2 $,& ~~$\Delta_{3}^N,~-\delta_3^N
  $   &~~$\Theta_{3},$& Mult. of SU(9)  \\[0.15em]
 \hline &&&&&&& && \\[-1.15em]
$\ominus=L$& $(---)$  &  $\frac{+5}{12}$ & $~~\frac{-3}{12},$& $\frac{0}{12}$ &$\frac{189}{144}  $  &$\frac{+243}{144}$ & $ \frac{0}{12},~~\frac{- 6}{12}$&$\frac{+8}{12} +\frac{+2}{12}  $& 1  \\ [0.1em]
 \hline &&&&&&&&& \\[-1.25em]
  $\oplus=R$& $(+++)$  &  $\frac{-5}{12}$ & $~~\frac{-3}{12}$&$\frac{0}{12}$  &$\frac{189}{144}$ &$\frac{+243}{144}$&$ \frac{0}{12},~~\frac{-6}{12}$& $\frac{-2}{12}+\frac{2}{12} $  &2   \\[0.15em] 
\hline
\end{tabular}
\end{center}
\caption{Two indices spinor-form from $T_9^0$:   Thus, we obtain $ {\color{red}(\tsix,{\one})_R}+
(\tsix,{\one})_L+(\tsix,{\one})_R $. }\label{tab:AbbTnine} 
\end{table}

 \bigskip
\noindent$\bullet$   One index spinor-form from $T_9^0$:
 the spinor forms satisfying  $(P+9V_0)^2=\frac{234}{144}=\frac{13}{8}$  and $12(P+9V_0)\cdot a_3= 0$ are possible for SU(9):
  \begin{equation}
 P_{9}=(\underline{+\,-^{8}};\frac{-3}{2},-;-----),\label{eq:NinebAbar}
 \end{equation}
 for which the massless fields are are projected out since $\Theta_9$ in the top row is $\frac{-1}{12}$.   
   
\subsection{Twisted sector $T_4\,  ( \delta_4 =0)$}

In $T_4$,   $4V_{0,+,-}$ become
\begin{eqnarray}
4V_0&=& \left( (\frac{+4}{12})^{9};1,\,2 \,;\,(2)^5\,\right), \\
4V_+&=& \left((\frac{+4}{12})^{9};1,\,\frac{+8}{12}\,;(\frac{+40}{12})^5\,\right), \\
4V_-&=& \left((\frac{+4}{12})^{9};1,\,\frac{+40}{12}\,;(\frac{+8}{12})^5\,\right) .  
\end{eqnarray}

\noindent$\bullet$  One index vector-form  from $T_4^0$:   
\begin{eqnarray}
P_9= (\underline{-1~0^{8}}\,;-1~-2;(-2)^5). 
\end{eqnarray}
satisfies $(P_9+4V_0)^2=\frac{192}{144}$.   
 Thus, we obtain Table \ref{tab:OneFourZero}.
 
 \begin{table}[!h]
\begin{center}
\begin{tabular}{|cc|c|cc|ccc|cc|}
 \hline &&&&&&&&& \\[-1.15em]
  Chirality  &   $\tilde s$& $-\tilde{s}\cdot\phi_s$& $-k\,p_{\rm vec}^{ k\,\rm th}\cdot \phi_s$ &$k\,P_{9}\cdot V_+$& $(k/2)\phi_s^2$,&$  -(k/2) V_+^2 $,& ~~$\Delta_{4}^N,~-\delta_4^N
  $   &~~$\Theta_{4},$& Mult. of SU(9)  \\[0.15em]
 \hline &&&&&&& && \\[-1.15em]
$\ominus=L$& $(---)$  &  $\frac{+5}{12}$ & $~~\frac{+8}{12},$& $\frac{-4}{12}$ &$\frac{84}{144} $  &$\frac{+108}{144}$ & $ \frac{0}{12},~~\frac{0}{12}$&$\frac{+9}{12}  $& 7 \\ [0.1em]
 \hline &&&&&&&&& \\[-1.25em]
  $\oplus=R$& $(+++)$  &  $\frac{-5}{12}$ & $~~\frac{+8}{12}$&$\frac{+4}{12}$ &$\frac{84}{144} $  &$\frac{+108}{144}$ & $ \frac{0}{12},~~\frac{0}{12}$&$\frac{-1}{12}  $&  0 \\[0.15em] 
\hline
\end{tabular}
\end{center}
\caption{One index vector-form from $T_4^0$:  Thus,  we obtain ${\color{red} 7({\nineb},{\one})_L} $. } \label{tab:OneFourZero}    
\end{table}
 
\noindent$\bullet$  One index spinor-form  from $T_4^+$:   
\begin{equation}
P_9=\left( \underline{+\,-^8};-,\,-\,;\, {(\frac{-7}{2})^5}\right). 
\label{eq:VfourPlus}
\end{equation}
satisfies $(P_9+4V_+)^2=\frac{192}{144}$.   
 Thus, we obtain Table \ref{tab:OneFourPlus}.
 
 \begin{table}[!h]
\begin{center}
\begin{tabular}{|cc|c|cc|ccc|cc|}
 \hline &&&&&&&&& \\[-1.15em]
  Chirality  &   $\tilde s$& $-\tilde{s}\cdot\phi_s$& $-k\,p_{\rm vec}^{ k\,\rm th}\cdot \phi_s$ &$k\,P_{9}\cdot V_+$& $(k/2)\phi_s^2$,&$  -(k/2) V_+^2 $,& ~~$\Delta_{4}^N,~-\delta_4^N
  $   &~~$\Theta_{4},$& Mult. of SU(9)  \\[0.15em]
 \hline &&&&&&& && \\[-1.15em]
$\ominus=L$& $(---)$  &  $\frac{+5}{12}$ & $~~\frac{+8}{12},$& $\frac{0}{12}$ &$\frac{84}{144} $  &$\frac{+108}{144}$ & $ \frac{0}{12},~~\frac{0}{12}$&$\frac{+1}{12}  $& 0 \\ [0.1em]
 \hline &&&&&&&&& \\[-1.25em]
  $\oplus=R$& $(+++)$  &  $\frac{-5}{12}$ & $~~\frac{+8}{12}$&$\frac{+4}{12}$ &$\frac{84}{144} $  &$\frac{+108}{144}$ & $ \frac{0}{12},~~\frac{0}{12}$&$\frac{-9}{12}  $&  7 \\[0.15em] 
\hline
\end{tabular}
\end{center}
\caption{One index spinor-form from $T_4^+$:   We obtain ${\color{red} 7({\nineb},{\one})_R} $.  } \label{tab:OneFourPlus}    
\end{table}

   \bigskip
 \noindent$\bullet$ One index spinor-form for $T_4^-$:  the spinor form satisfying  $(P+4V_-)^2=\frac{192}{144}$  and $12(P+4V_0)\cdot a_3= 0$ are possible  for SU(9):
\begin{equation}
P_9=\left( \underline{+\,-^8};-,\,\frac{-7}{2}\,;\, {-^5}\right). 
\label{eq:Vfour2}
\end{equation}
 The massless specta are presented in  
Table \ref{tab:Sin4Minus}. 

 \begin{table}[!ht]
\begin{center}
\begin{tabular}{|cc|c|cc|ccc|cc|}
 \hline &&&&&&&&& \\[-1.15em]
  Chirality  &   $\tilde s$& $-\tilde{s}\cdot\phi_s$& $-k\,p_{\rm vec}^{ k\,\rm th}\cdot \phi_s$ &$k\,P \cdot V_-$& $(k/2)\phi_s^2$,&$  -(k/2) V_-^2 $,& ~~$\Delta_{4}^N,~-\delta_4^N
  $   &~~$\Theta_{4},$& Singlets \\[0.15em]
 \hline &&&&&&& && \\[-1.15em]
$\ominus=L$& $(---)$  &  $\frac{+5}{12}$ & $~~\frac{+8}{12},$& $\frac{0}{12}$ &$\frac{84}{144} $  &$\frac{-276}{144}$ & $ \frac{0}{12},~~\frac{0}{12}$&$\frac{+1}{12}  $& 0  \\ [0.1em]
 \hline &&&&&&&&& \\[-1.25em]
  $\oplus=R$& $(+++)$  &  $\frac{-5}{12}$ & $~~\frac{+8}{12}$&$\frac{0}{12}$ &$\frac{84}{144} $  &$\frac{-276}{144}$ & $ \frac{0}{12},~~\frac{0}{12}$&$\frac{-9}{12}  $&  7  \\[0.15em] 
\hline
\end{tabular}
\end{center}
\caption{One index spinor-form from $T_4^-$:  We obtain ${\color{red} 7({\nineb},{\one})_R} $.   } \label{tab:Sin4Minus}  
\end{table}

 \subsection{Twisted sector $T_1 \,( \delta_1 =\frac{1}{12})$}
 \noindent
In $T_1$,  we use Eq. (\ref{eq:Model}). 
 
 \bigskip
\noindent$\bullet$  One index vector-form for $T_1^+$: the vector 
\begin{eqnarray}
P_{5}=(0^9;0,0;\underline{-1,\,-1,\,-1,\,-1,\,0 }), 
\end{eqnarray}
satisfies $(P+V_+)^2=\frac{138}{144}$   which is short by $\frac{72}{144}$ from the target value of  $\frac{210}{144}$, but the massless fields are projected out.   

\bigskip
\noindent
On the other hand, the lattice point
\begin{eqnarray}
P_{9}=(\underline{-1~0^8};0,0;-1,\,-1,\,-1,\,-1,\,-1), 
\end{eqnarray}
satisfies $(P+V_+)^2=\frac{162}{144}$   which is short by $\frac{48}{144}=\frac{4}{12}$ from the target value of  $\frac{210}{144}$, and the spectra are shown in Table \ref{tab:T1VectorPlus}.  

\begin{table}[!ht]
\begin{center}
\begin{tabular}{|cc|c|cc|ccc|cc|}
 \hline &&&&&&&&& \\[-1.15em]
  Chirality  &   $\tilde s$& $-\tilde{s}\cdot\phi_s$& $-k\,p_{\rm vec}^{ k\,\rm th}\cdot \phi_s$ &$k\,P_{5}\cdot V_+$& $(k/2)\phi_s^2$,&$  -(k/2) V_+^2 $,& ~~$\Delta_{k}^N,~-\delta_k^N
  $   &~~$\Theta,$& Mult. of SU(5)$'$  \\[0.15em]
 \hline &&&&&&& && \\[-1.15em]
$\ominus=L$& $(---)$  &  $\frac{+5}{12}$ & $~~\frac{+5}{12},$& $\frac{-3}{12}$ &$\frac{+21}{144} $  &$\frac{+27}{144}$ & $ \frac{+4}{12},~~\frac{-2}{12}$&$\frac{+1}{12}  $& 0 \\ [0.1em]
 \hline &&&&&&&&& \\[-1.25em]
  $\oplus=L$& $(+++)$  &  $\frac{-5}{12}$ & $~~\frac{+5}{12}$&$\frac{-4}{12}$ &$\frac{+21}{144} $  &$\frac{+27}{144}$ & $ \frac{+2}{12},~~\frac{-2}{12}$&$\frac{-9}{12}  $&  3  \\[0.15em] 
\hline
\end{tabular}
\end{center}
\caption{One index spinor-form from $T_1^+$:   Thus, we obtain ${\color{red}3({\nine},{\one})_R} $.  } \label{tab:T1VectorPlus}
\end{table}

 \bigskip
\noindent$\bullet$  One index vector-form for $T_1^-$: the vector 
\begin{eqnarray}
P_{9}=(\underline{-1~0^8};0,-1;0^5), 
\end{eqnarray}
satisfies $(P+V_-)^2=\frac{162}{144}$   which is short by $\frac{48}{144}=\frac{4}{12}$ from the target value of  $\frac{210}{144}$, and the spectra are shown in Table \ref{tab:NineOneMin} .  

\begin{table}[!ht]
\begin{center}
\begin{tabular}{|cc|c|cc|ccc|cc|}
 \hline &&&&&&&&& \\[-1.15em]
  Chirality  &   $\tilde s$& $-\tilde{s}\cdot\phi_s$& $-k\,p_{\rm vec}^{ k\,\rm th}\cdot \phi_s$ &$k\,P_{9}\cdot V_+$& $(k/2)\phi_s^2$,&$  -(k/2) V_+^2 $,& ~~$\Delta_{1}^N,~-\delta_1^N
  $   &~~$\Theta_{1},$& Mult. of SU(9)  \\[0.15em]
 \hline &&&&&&& && \\[-1.15em]
$\ominus=L$& $(---)$  &  $\frac{+5}{12}$ & $~~\frac{+5}{12},$& $\frac{-11}{12}$ &$\frac{+21}{144} $  &$\frac{-69}{144}$ & $ \frac{+4}{12},~~\frac{-2}{12}$&$\frac{-3}{12}  $& 3 \\ [0.1em]
 \hline &&&&&&&&& \\[-1.25em]
  $\oplus=L$& $(+++)$  &  $\frac{-5}{12}$ & $~~\frac{+5}{12}$&$\frac{-11}{12}$ &$\frac{+21}{144} $  &$\frac{-69}{144}$ & $ \frac{+4}{12},~~\frac{-2}{12}$&$\frac{-1}{12}  $&  0  \\[0.15em] 
\hline
\end{tabular}
\end{center}
\caption{One index vector-form from $T_1^-$:  Thus, we obtain ${\color{red}3\cdot ({\nine},{\one})_L} $.}\label{tab:NineOneMin} 
\end{table}

\subsection{Twisted sector $T_2 \,( \delta_2 =\frac{0}{12} )$}
 
In $T_2$, $2V_{0,+,-}$ are
\begin{eqnarray}
2V_0&=&\left( (\frac{+2}{12})^{9};\,\frac{+6}{12},\,\frac{+12}{12};\, (\frac{+12}{12})^5  \right), \\   
2V_+&=&\left( (\frac{+2}{12})^{9};\,\frac{+6}{12},\,\frac{+4}{12};\, (\frac{+20}{12})^5  \right), \\ 
2V_- &=& \left( (\frac{+2}{12})^{9};\,\frac{+6}{12},\,\frac{+20}{12};\, (\frac{+4}{12})^5 \right).\label{eq:2VMinus}
\end{eqnarray}
 
\noindent$\bullet$  One index vector-form from $T_2^0$:
 the vector 
\begin{eqnarray}
P_{9}=(0^9;-1,-1;\underline{ -1,-1,-1,-1,0}), 
\end{eqnarray}
satisfies $(P+2V_0)^2=\frac{216}{144}$. Massless fields are shown in  Table \ref{tab:TtwoZeroMinus}. 
\begin{table}[!ht]
\begin{center}
\begin{tabular}{|cc|c|cc|ccc|cc|}
 \hline &&&&&&&&& \\[-1.15em]
  Chirality  &   $\tilde s$& $-\tilde{s}\cdot\phi_s$& $-k\,p_{\rm vec}^{ k\,\rm th}\cdot \phi_s$ &$k\,P \cdot V_0$& $(k/2)\phi_s^2$,&$  -(k/2) V_0^2 $,& ~~$\Delta_{1}^N,~-\delta_1^N
  $   &~~$\Theta_{1},$& Singlets \\[0.15em]
 \hline &&&&&&& && \\[-1.15em]
$\ominus=L$& $(---)$  &  $\frac{+5}{12}$ & $~~\frac{+5}{12},$& $\frac{-6}{12}$ &$\frac{+42}{144} $  &$\frac{+54}{144}$ & $ \frac{0}{12},~~\frac{0}{12}$&$\frac{+4}{12}  $& 0 \\ [0.1em]
 \hline &&&&&&&&& \\[-1.25em]
  $\oplus=L$& $(+++)$  &  $\frac{-5}{12}$ & $~~\frac{+5}{12}$&$\frac{-6}{12}$ &$\frac{+42}{144} $  &$\frac{+54}{144}$ & $ \frac{0}{12},~~\frac{0}{12}$&$\frac{-6}{12}  $&  3  \\[0.15em] 
\hline
\end{tabular}
\end{center}
\caption{One index vecror-form from $T_2^0$:  Thus, we obtain ${\color{red}3\cdot ({\one},{\five}')_R} $.} \label{tab:TtwoZeroMinus} 
\end{table}

\bigskip
\noindent$\bullet$ One index vector-form from $T_2^+$:
 the vector 
\begin{eqnarray}
P_{5}=(0^9;-1,0;\underline{ -1,-2,-2,-2,-2}), 
\end{eqnarray}
satisfies $(P+2V_+)^2=\frac{216}{144}$, and  masless fields are projected out.
 

\bigskip
\noindent$\bullet$  One index vector-form from $T_2^-$:
 the vector 
\begin{eqnarray}
P_{5}=(0^9;-1,-2;\underline{ -1,0,0,0,0}), 
\end{eqnarray}
satisfies $(P+2V_-)^2=\frac{216}{144}$, which gives Table \ref{tab:TtMinusOne} . 

\begin{table}[!ht]
\begin{center}
\begin{tabular}{|cc|c|cc|ccc|cc|}
 \hline &&&&&&&&& \\[-1.15em]
  Chirality  &   $\tilde s$& $-\tilde{s}\cdot\phi_s$& $-k\,p_{\rm vec}^{ k\,\rm th}\cdot \phi_s$ &$k\,P_{5}\cdot V_-$& $(k/2)\phi_s^2$,&$  -(k/2) V_-^2 $,& ~~$\Delta_{2}^N,~-\delta_2^N
  $   &~~$\Theta_{2},$& Mult. of SU(9)  \\[0.15em]
 \hline &&&&&&& && \\[-1.15em]
$\ominus=L$& $(---)$  &  $\frac{+5}{12}$ & $~~\frac{+10}{12},$& $\frac{-2}{12}$ &$\frac{+42}{144} $  &$\frac{+54}{144}$ & $ \frac{0}{12},~~\frac{0}{12}$&$\frac{+1}{12}  $& 0 \\ [0.1em]
 \hline &&&&&&&&& \\[-1.25em]
  $\oplus=L$& $(+++)$  &  $\frac{-5}{12}$ & $~~\frac{+10}{12}$&$\frac{-2}{12}$ &$\frac{+42}{144} $  &$\frac{+54}{144}$ & $ \frac{0}{12},~~\frac{0}{12}$&$\frac{-9}{12}  $&  3  \\[0.15em] 
\hline
\end{tabular}
\end{center}
\caption{One index vector-form from $T_2^-$:  Thus, we obtain ${\color{red}3\cdot ({\one},{\fiveb})_R} $.}\label{tab:TtMinusOne} 
\end{table}
 
\subsection{Twisted sector $T_7 \,( \delta_7 =\frac{1}{12} )$}
 
For $T_7^{0,+,-}$, we have
\begin{eqnarray}
7V_0&=&\left( (\frac{+7}{12})^{9};\,\frac{+21}{12},\,\frac{+42}{12};\, (\frac{+42}{12})^5 \right), \\   
7V_+&=&\left( (\frac{+7}{12})^{9};\,\frac{+21}{12},\,\frac{+14}{12};\, (\frac{+70}{12})^5 \right),\\
7V_- &=& \left( (\frac{+7}{12})^{9};\,\frac{+21}{12},\,\frac{+70}{12};\, (\frac{+14}{12})^5  \right).
\end{eqnarray}
  
\noindent$\bullet$  One index spinor-form from $T_7^0$: 
\begin{eqnarray}
P_{9}=\left(\underline{\frac{-3}{2}\,-^{8}};\frac{-3}{2},\frac{-7}{2};(\frac{-7}{2})^5\right), \label{eq:V7Zero2s
}
\end{eqnarray}
gives $(P_9+7V_0)^2=  \frac{138}{144}$, which is short of $\frac{6}{12}$ from $\frac{210}{12}$, and we obtain Table  XXI.

\begin{table}[!ht]
\begin{center}
\begin{tabular}{|cc|c|cc|ccc|cc|}
 \hline &&&&&&&&& \\[-1.15em]
  Chirality  &   $\tilde s$& $-\tilde{s}\cdot\phi_s$& $-k\,p_{\rm vec}^{ k\,\rm th}\cdot \phi_s$ &$k\,P_{9}\cdot V_0$& $(k/2)\phi_s^2$,&$  -(k/2) V_0^2 $,& ~~$\Delta_{k}^N,~-\delta_k^N
  $   &~~$\Theta_{9},$& Mult. of SU(9)  \\[0.15em]
 \hline &&&&&&& && \\[-1.15em]
$\ominus=L$& $(---)$  &  $\frac{+5}{12}$ & $~~\frac{+19}{12},$& $\frac{-4}{12}$ &$\frac{+147}{144} $  &$\frac{+189}{144}$ & $ \frac{+6}{12},~~\frac{-2}{12}$&$\frac{+4}{12}  $& 0 \\ [0.1em]
 \hline &&&&&&&&& \\[-1.25em]
  $\oplus=L$& $(+++)$  &  $\frac{-5}{12}$ & $~~\frac{+19}{12}$&$\frac{-4}{12}$ &$\frac{+147}{144} $  &$\frac{+189}{144}$ & $ \frac{+6}{12},~~\frac{-2}{12}$&$\frac{-6}{12}  $&  3 \\[0.15em] 
\hline
\end{tabular}
\end{center}
\caption{One index spinor-form from $T_7^0$: $\color{red} 3\,({\nine},{\one})_R  $.}  
\end{table}

 \bigskip
\noindent$\bullet$  Two index spinor-form from $T_7^-$:
\begin{eqnarray}
P_{5}=\left(-^9;\frac{-3}{2},\frac{-11}{2};\underline{(\frac{-1}{2})^2\,(\frac{-3}{2})^3} \right),  \label{eq:T5Minus1}
\end{eqnarray}
gives $(P_5+7V_-)^2=  \frac{210}{144}$, and massless fields are projected out.

 \bigskip
\noindent$\bullet$  One index spinor-form from $T_7^-$:
\begin{eqnarray}
P_{5}=\left(-^9;\frac{-3}{2},\frac{-13}{2};\underline{(\frac{-1}{2})\,(\frac{-3}{2})^4} \right),  \label{eq:T5Minus1}
\end{eqnarray}
gives $(P_5+7V_-)^2=  \frac{210}{144}$, and massless fields are projected out.

\subsection{Twisted sector $T_6 \,( \delta_6 =\frac{0}{12} )$}
\label{Ssec:Tsix}

\noindent
Twisted sectors $T_6^{0,+,-}$  are not distinguish by Wilson lines. So, in $T_6$ we just calculate the spectra whose multiplicity should be 3.
\begin{eqnarray}
  \color{red}
&&6V_0=\left( (\frac{+6}{12})^9  {\color{blue}\,;\,}\frac{+18}{12},\,  3 {\color{blue}\,;\,} 3^5 \right) .    
\end{eqnarray}

 \bigskip
\noindent$\bullet$   Two indices spinor-form from $T_6^0$: The spinor with the even number of +'s\footnote{The + and -   represent $\frac{+1}{2}$ and  $\frac{-1}{2}$, respectively.}
\begin{equation}
P_5=(-^9; \frac{-3}{2},\frac{-5}{2};\underline{(\frac{-5}{2})^2\,(\frac{-7}{2})^3}) \label{eq:ten}
\end{equation}
satisfies $(P_5+6V_0)^2=\frac{216}{144} $  which saturates the masslessness condition, and we obtain Table  \ref{tab:TensChiral6} with an extra angle $\frac{2}{12}$. 

 \begin{table}[!ht]
\begin{center}
\begin{tabular}{|cc|c|c|ccc|cc|}
 \hline &&&&&&&& \\[-1.15em]
  Chirality  &   $\tilde s$& $-\tilde{s}\cdot\phi_s$& $-p_{\rm vec}^{ k\,\rm th}\cdot \phi_s,k\,P_{5}\cdot V_0$& $(k/2)\phi_s^2$,&$  -(k/2) V_0^2 $,& ~~$\Delta_{1}^{N},~-\delta_2^N
  $   &$\Theta_5,$& Mult. of SU(5)  \\[0.15em]
 \hline &&&&&&& & \\[-1.15em]
$\ominus=L$& $(---)$  &  $\frac{+5}{12}$ & $~~\frac{0}{12},~~~~~~\frac{+6}{12}$ &$\frac{+126}{144}  $  &$\frac{+162}{144}$ & $ \frac{0}{12},~~\frac{0}{12}$&$\frac{-1}{12}+\frac{2}{12} $& 4\\ [0.1em]
 \hline &&&&&&&& \\[-1.25em]
$\oplus=L$& $(+++)$  &  $\frac{-5}{12}$ & $~~\frac{0}{12},~~~~~~\frac{+6}{12}$ &$\frac{+126}{144} $  &$\frac{+162}{144}$ & $ \frac{0}{12},~~\frac{0}{12}$&$\frac{-11}{12}+\frac{2}{12}$&  5  \\[0.15em] 
\hline
\end{tabular}
\end{center}
\caption{Two indices   spinor-forms from $V_6^0$ with an extra phase $e^{2\pi i\,\frac{+2}{12}}$. Originally, before providing the vacuum phase, there were four vectorlike pairs, and hence we obtain
${ {\color{red}3\cdot ({\one},{\tenb}')_R}
+24\cdot[({\one}, {\ten}')_L+({\one}, {\ten}')_R]}$.  }\label{tab:TensChiral6}  
\end{table}
   
 \bigskip
\noindent$\bullet$  One index spinor-form from $T_6^0$: The spinor-form \begin{equation}
P_5=(-^9; \frac{-3}{2},\frac{-7}{2};\underline{(\frac{-5}{2})\,(\frac{-7}{2})^4})  \label{eq:five}
\end{equation}
saturates the masslessness condition, and we obtain Table  \ref{tab:FivsChiral6}.  

 \begin{table}[!ht]
\begin{center}
\begin{tabular}{|cc|c|c|ccc|cc|}
 \hline &&&&&&&& \\[-1.15em]
  Chirality  &   $\tilde s$& $-\tilde{s}\cdot\phi_s$& $-p_{\rm vec}^{ k\,\rm th}\cdot \phi_s,k\,P_{5}\cdot V_0$& $(k/2)\phi_s^2$,&$  -(k/2) V_0^2 $,& ~~$\Delta_{1}^{N},~-\delta_2^N
  $   &$\Theta_5,$& Mult. of SU(5)  \\[0.15em]
 \hline &&&&&&& & \\[-1.15em]
$\ominus=L$& $(---)$  &  $\frac{+5}{12}$ & $~~\frac{0}{12},~~~~~~\frac{+6}{12}$ &$\frac{+126}{144}  $  &$\frac{+162}{144}$ & $ \frac{0}{12},~~\frac{0}{12}$&$\frac{-1}{12}+\frac{2}{12} $& 4\\ [0.1em]
 \hline &&&&&&&& \\[-1.25em]
$\oplus=L$& $(+++)$  &  $\frac{-5}{12}$ & $~~\frac{0}{12},~~~~~~\frac{+6}{12}$ &$\frac{+126}{144} $  &$\frac{+162}{144}$ & $ \frac{0}{12},~~\frac{0}{12}$&$\frac{-11}{12}+\frac{2}{12}$&  5  \\[0.15em] 
\hline
\end{tabular}
\end{center}
\caption{One index spinor-forms from $V_6^0$ with an extra phase $e^{2\pi i\,\frac{+2}{12}}$. Originally, there were four vectorlike pairs, and we obtain
${ {\color{red}3\cdot ({\one},{\five}')_R}
+24\cdot[({\one}, {\five}')_L+({\one}, {\five}')_R]}$.  }\label{tab:FivsChiral6}  
\end{table}

\bigskip
\noindent
Summarizing the non-singlet chiral fields 
we obtained,
\begin{eqnarray}
\color{red} \nineb_R(U)+\tsix_R(T_9^0)+7\cdot \nineb_L(T_4^0)  +7\cdot \nineb_R(T_4^+)  +7\cdot \nineb_R(T_4^-)  +  3\cdot \nine_R(T_1^+)+ 3\cdot \nine_L(T_1^-)+ 3\cdot \nine_R(T_7^0), \label{eq:FNine}
 \end{eqnarray}
 that lead to $\tsix_R+5\cdot \nineb_R$, and 
 \begin{eqnarray}
\color{red}  3\cdot  {\five}'_R(T_2^0) +3\cdot  {\fiveb}'_R(T_2^-)  + 3\cdot {\tenb}'_R(T_6)+3\cdot {\five}'_R(T_6).\label{eq:FFive}
\end{eqnarray}
 These spectra in Eqs. (\ref{eq:FNine}) and (\ref{eq:FFive}) do not lead to non-Abelian gauge anomalies.
This is another proof of providing the vacuum phase from the anomaly constraint.
    
\section{Singlets}\label{sec:Singlets} 

\noindent Here, we summarize chiral singlets in Table \ref{tab:ChiralSinglets}.  We also show  $({\ten}'_L+{\ten}'_R)$'s,  which we obtained in Subsect. \ref{Ssec:Tsix}, which can acquire GUT scale VEVs.

 \begin{table}[!ht]
\begin{center}
\begin{tabular}{|c|c|c|c|c|}
 \hline &&&&  \\[-1.15em]
  Twisted Sector &   $P~;~P+kV_i$& $-\hat{s}\cdot\phi_s$
    &   $\Theta_k$ &${\cal P}_k$ (L or R)
  \\
 \hline &&&& \\[-1.4em]
$T_3^0$& $\left(-^9;-,\frac{-3}{2};(\frac{-3}{2})^5\right);\left((\frac{-3}{12})^9;\frac{+3}{12},0;\,0^5\right )$  & $\frac{5}{12}$& $\frac{+5}{12}$& 3(L),~3(R) \\ 
 &&&&  \\[-1.25em]
$T_3^+$& $\left(-^9;-,-;(\frac{-5}{2})^5\right);\left((\frac{-3}{12})^9;\frac{+3}{12},0;\,0^5\right )$ & $\frac{5}{12}$  &  $\frac{+5}{12}$& 3(L),~3(R)  \\[0.15em] 
 &&&& \\[-1.25em]
$T_3^-$& $\left(-^9;\frac{-3}{2},-;-^5\right);\left((\frac{-3}{12})^9;\frac{-9}{12},0;\,0^5\right )$ & $\frac{5}{12}$  &  $\frac{+5}{12}$& 3(L),~3(R)  \\[0.15em] 
\hline &&&& \\[-1.15em]
$T_4^0$& $\left(-^9;-,\frac{-5}{2};(\frac{-5}{2})^5\right);\left((\frac{-2}{12})^9;\frac{+6}{12},\frac{-1}{12};\,(\frac{-1}{12})^5\right )$  & $\frac{5}{12}$ &  $\frac{+1}{12}$& 7(R) \\ [0.1em]
 && && \\[-1.25em]
$T_4^+$& $\left(-^9;\frac{-3}{2},-;(\frac{-7}{2})^5\right);\left((\frac{-2}{12})^9;\frac{-6}{12},\frac{+2}{12};\,(\frac{-2}{12})^5\right )$  & $\frac{5}{12}$ &  $\frac{+1}{12}$& 7(R) \\ [0.1em]
 && && \\[-1.25em]
$T_4^-$& $\left(-^9;\frac{-3}{2},\frac{-7}{2};-^5\right);\left((\frac{-2}{12})^9;\frac{-6}{12},\frac{-2}{12};\,(\frac{+2}{12})^5\right )$  & $\frac{5}{12}$ &  $\frac{+5}{12}$& 3(L),~3(R) \\ [0.1em]
\hline && && \\[-1.25em]
$T_1^+$& $\left(0^9;-1,0;(-1)^5\right);\left((\frac{+1}{12})^9;\frac{-9}{12},\frac{+2}{12};\,(\frac{-2}{12})^5\right )$  & $\frac{5}{12}$ &  $\frac{+9}{12}$& 3(L) \\ [0.15em]
  \hline  &&&&  \\[-1.25em]
 $T_2^-$& $\left(0^9;-1,-1;0^5\right);\left((\frac{+2}{12})^9;\frac{-6}{12},\frac{+8}{12};\,(\frac{+4}{12})^5\right )$  & $\frac{5}{12}$ &  $\frac{+1}{12}$& 3(R) \\ [0.15em]
  \hline  &&&&  \\[-1.25em]
$T_7^-$& $\left(-^9;\frac{-3}{2},\frac{-13}{2};(\frac{-3}{2})^5\right);\left((\frac{+1}{12})^9;\frac{+3}{12},\frac{-8}{12};\,(\frac{-4}{12})^5\right )$  & $\frac{5}{12}$ &  $\frac{+5}{12}$& 3(L),~3(R) \\ [0.25em]
\hline
$T_6^0$& $\left( -^9;\frac{-3}{2},\frac{-5}{2};(\frac{-7}{2})^5\right);\left( 0^9;0,\frac{+6}{2};(\frac{-6}{2})^5\right)$ & $\frac{5}{12}$  &  $\frac{+6}{12}$ &   6(L),~5(R)\\[0.25em] 
$T_6^0$& $\left( -^9;\frac{-3}{2},\frac{-7}{2};(\frac{-5}{2})^5\right);\left( 0^9;0,\frac{-6}{2};(\frac{+6}{2})^5\right)$ & $\frac{5}{12}$  &  $\frac{+6}{12}$ & 3$\times$6(L),~3$\times$5(R)\\[0.25em] 
\hline\hline
$T_6$& $\left(-^9;\frac{-3}{2},\frac{-5}{2};\underline{(\frac{-5}{2})^2 (\frac{-7}{2})^3}\right);\left(0^9;0,\frac{+6}{12};\,\underline{(\frac{+6}{2})^2 (\frac{-6}{2})^3}\right )$  & $\frac{5}{12}$ &  $\frac{+1}{12}$& 3$\times 8[{\ten}'_L+{\ten}'_R]$ \\ [0.35em]
\hline
\end{tabular}
\end{center}
\caption{Summary of chiral singlets, $\one$'s. The value for $-\hat{s}\cdot\phi_s$ is for $(\ominus;---)$, \ie the ones in the top rows in the tables. We also listed the vectorlike $({\one},{\ten}\oplus  {\tenb})_L $ obtained previously in $T_6$.}\label{tab:ChiralSinglets} 
\end{table}
\section{Three families of SU(5)$'$}\label{sec:Families}

\noindent
 Let us interpret the SU(5)$'$ GUT as the observable sector
with the three families from
 \begin{eqnarray}
 3\cdot {\tenb}'_R(T_6) + 3\cdot [{\five}'_R(T_2^0)+ {\fiveb}'_R(T_2^-)  +{\five}'_R(T_6)]\to 3\,\{{\ten}'_L +{\fiveb}'_{mL}\}\oplus {\rm vectorlike~pairs}.\label{eq:ThreeFam}
\end{eqnarray}
Vectorlike pair(s) of ${\five}'+{\fiveb}'$ in Eq. (\ref{eq:ThreeFam}) are $ {\fiveb}'_R(T_2^-)$'s and linear combinations of $\five$'s. Using the charge conjugated fields of (\ref{eq:ThreeFam}), let
\begin{equation}
H_5=\{\cos\theta\, {\five}'_L(T_2^0)+\sin\theta\, {\five}'_L(T_6),{\five}'_L(T_2^+)\},~~\overline{H}_{\bar 5}=\cos\theta\, {\fiveb}'_L(T_2^0)+\sin\theta\, {\fiveb}'_L(T_6)
\end{equation} 
be interprested as the Higgs quintets. Then, three matter quintets are 
\begin{equation}
{\fiveb}'_{mL}=-\sin\theta\, {\fiveb}'_L(T_2^0)+\cos\theta\, {\fiveb}'_L(T_6)
\end{equation}

\bigskip
\noindent
Now, let us comment on the largest and the smallest mass scales in the SM.

\subsection{Top mass}

\noindent The top quark mass arises, e.g. for $\theta=0$, from 
 \begin{eqnarray}
 T_6({\ten}'_L) T_6({\fiveb}'_{mL})T_2^0(\overline{H}_{{\bar 5}'})T_{10}({\one}_L) \label{eq:YukDorm}
 \end{eqnarray}
 where $\langle {\one}_L\rangle$ is of order the GUT scale. ${\one}_L$ is from $T_2^-$ of \ref{tab:ChiralSinglets} whose Wilson line component -- is not cancelled by $T_6$ and $T_2^0$. Therefore, we should take $\theta\ne 0$. Wilson lines must match appropriately for (\ref{eq:YukDorm}) to be present. The L-handed top quark is taken as the permutation singlet
 \begin{eqnarray}
 t_L\sim \frac{1}{\sqrt3}({\ten}'_1+{\ten}'_2+{\ten}'_3)_L\label{eq:Ltop}
 \end{eqnarray}
 where three ${\ten}'$ imply the $\Qem=\frac{2}{3}$ quark in the quark doublet in  ${\ten}'$s(with the charge conjugated fields of  Eq. (\ref{eq:ThreeFam})). L-handed charm and up quarks are approximately orthogonal to (\ref{eq:Ltop}). Other masses in the SM fields can be studied similarly.
  
\subsection{Neutrino masses}
 
\bigskip
\noindent  Here, we comment on the neutrino masses. The effective Weinberg operator \cite{Weinberg79}  in the anti-SU(5) GUT is  
\begin{eqnarray}
 \overline{\five}'_{mL}\, \overline{\five}'_{mL} H_5\,H_5\label{eq:NeutMassW}
 \end{eqnarray}
If the mixing angle in the Higgs quintet is zero, then we expect the operator
\begin{eqnarray}
 \overline{\five}'_{mL}(T_6)\, \overline{\five}'_{mL}(T_6)\,   H_5(T_2)\,H_5(T_2)\,{\one}_L(T_8),\label{eq:NeutMassFlip}
 \end{eqnarray}
and the order of $m_\nu$ scale is
\begin{eqnarray}
 m_\nu\sim \left(\frac{\langle {\one}_L(T_8)\rangle}{\Mg} \right) \frac{v_{\rm ew}^2}{\Mg}\sim 10^{-3}\varepsilon\,{\eV},\label{eq:NtMassOrder}
 \end{eqnarray}
where $\varepsilon\sim \langle {\one}_L(T_8)\rangle/\Mg$. Of course, $\theta\ne 0$ is assumed to have non-vanishing contributions.

\section{Symmetry breaking}\label{sec:Spontaneous}

\noindent
Breaking GUT groups are broadly divided into two classes. These are done by two indices tensors, $\Phi^a_b(\equiv $ adjoint representation) \cite{GG74} and  $\Phi^{ab}\oplus \Phi_{ab}(\equiv $anti-symmetric representation given in Eq. (4)) \cite{DKN84}. In string compactification, at the level-1 construction, there is no adjoint representation.\footnote{For the rank 4 GUT SU(5), the F-theory introduces an adjoint representation, which is not arising from a ten dimensional string theory. For SU(9), it is impossible to obtain an adjoint representation.}  In our construction, there are U(1)$^4$ symmetry out of which we can pick up an appropriate U(1)$_X$ for the flipped-SU(5)/anti-SU(5) or anti-SU($N$). This choice of  U(1)$_X$ for $N\ge 5$ allows the \Uoneem~preserving VEV direction $\alpha_{45}$ and $-\alpha_{14}$ of Eq. (4) which can separate color and the rest.  Since we interpret SU(5)$'$ as the visible sector, SU(9) is used for triggering dynamical breaking of SUSY.
 
\bigskip
\noindent
For SUSY breaking, chiral spectra is needed.\footnote{Even without chiral spectra, SUSY breaking is possible with an extra confining force if the intervention of gravity is considered \cite{Witten81,Nilles82}. However, we disregard the help from gravity here.}  
Several decades ago, one family SU(5) was hinted for breaking SUSY \cite{Veneziano82}, and it was shown that the idea can be realized in a model from string compactification \cite{KimKyae19}. The confining force SU(9) can form the   condensates below the GUT scale when the SU(9) coupling constant becomes O(1),
\begin{equation}
S^{ij}\equiv \Psi^{AB}\Psi^i_{A}\Psi^j_{B}, \label{eq:ConfSinglet}
\end{equation}
where $ \Psi^{AB}$ is the chiral $\tsix$ of Table \ref{tab:AbbTnine}  and  $ \Psi^i_A$ are the remaining $\nineb$'s from Eq. (\ref{eq:FNine}),  after removing vector-like pairs.
With the appropriate choice of U(1)$_R$ charges for $\Psi^{AB}$ and $\Psi^i_{A}$ as done in Ref. \cite{KimKyae19}, $S^{ij}$ is 
restricted to carry two units of  U(1)$_R$ charge. Below the confining scale, therefore, the leading term in the superpotential is linear in $S$.
Since there remain five  $\nineb$'s, there are 10 independent SU(9) singlets formed below the SU(9) confining scale. Since we consider SUSY, we can construct a superptential in terms of some $S^{ij}$(where $ij$ is counting the number of singlets) below the confining scale $M_c$ as
\begin{equation}
W\sim 
M_c^2 S + M_c' S S'+\cdots \label{eq:WtwoTerms}
\end{equation}
Since we have the effective term above the confining scale, with an O(1) coupling, 
\begin{equation}
W_0\sim \Psi^{AB}\Psi^i_{A}\Psi^j_{B}\to M^2 S\label{eq:WoneTerm}
\end{equation} 
where $S$ is defined at the scale $M$. Comparing two terms in Eq. (\ref{eq:WtwoTerms}), we have $M_c\sim M$. For $M_c'$ of Eq. (\ref{eq:WtwoTerms}), we have $M_c'\sim M^4/\Mp^3$ which is $M^3/\Mp^3$ factor smaller than $M_c$, where $\Mp$ is the Planck mass and $M_c'$ is taken as the largest scale breaking U(1)$_R$. Below the GUT scale, there are complications due to the GUT symmetry breaking. So, if we take $M_c$ somewhat below the GUT scale, $M_c'$ is at least $10^{-6} ( \sim\Mg^3/\Mp^3$) times smaller than $M_c$. In this approxination, let us consider $W\sim M_c^2S$. This does not satisfy the SUSY condition: $\partial W/\partial S=M_c^2\ne 0$ where $M_c$ is considered to be nonzero because it was given by Eq. (\ref{eq:ConfSinglet}) \cite{KimKyae19}.

\bigskip
\noindent
The SUSY breaking discussed in this section needs a qualification in string compactification. The essential point is the appearance of chiral spectra containing two indices representation $\Psi^{[AB]}$. But, a chiral spectra containing  $\Psi^{[AB]}$ in SU($N$) with $N\ge 5$ from level-1 construction,  in addition to three visible sector families, was appeared previously only in Ref. \cite{Huh09}. The present model (\ref{eq:Model}) is the second example even with $N$ as large as 9. All satandard-like models so far considered have not addressed this question.  
 
\section{Conclusion}\label{sec:Conclusion}

\noindent
In this paper, compactifying the SO(32) heterotic string with 
$\Z_{12-I}$ orbifold symmetry, we obtained one family SU(9) GUT  and three families SU(5)$'$ GUT.  Sect. \ref{section:Orb} is a brief summary of compactification schemes presented in \cite{LNP954}.  In this paper, we realized chiral representations: $\tsix\oplus 5\cdot\nineb $ for a SU(9) GUT and  $3\{{\ten}'_L\oplus {\fiveb}'_L\}$ for a SU(5)$'$ GUT.  These chiral spectra  do not lead to non-Abelian gauge anomalies.   The anti-SU($N$) presented in this paper is a completely different class from the flipped-SU($N$)s from the spinor representations of SO($2N$). The visible sector is interpreted by the SU(5)$'$ GUT with three families $3\cdot (\ten'\oplus \fiveb')$, among which the L-handed top quark is interpreted as the most symmetric combination, for example by $t_L\sim \frac{1}{\sqrt3}({\ten}_1'+{\ten}_2'+{\ten}_3')_L$. SUSY breaking is dynamically achieved by the chiral specra of SU(9) GUT.
  
 \bigskip
 \noindent We presented in some detail  the Yukawa couplings for the heaviest fermion, $t$, and also for the neutrinos. Other fermion masses can be obtained similarly.
Spontaneous symmetry breaking of the visible sector SU(5)$'$ GUT  is achieved  by Higgsing of the anti-symmetric tensor representations, $\langle \ten'\rangle \oplus\langle \tenb'\rangle $.   In particular, we confirm that the non-Abelian anomalies of SU(9) gauge group vanish and hence our compactification scheme achieves the key requirement of the 4D effective field theory. In the supersymmetric version, we presented a scenario how supersymmetry can be broken dynamically via the confining gauge group SU(9). So far, most phenomenological studies from string compactification relied on $\EE8$ heterotic string, and this invites the SO(32) heterotic string very useful for future phenomenological
studies.

 \bigskip
 \noindent  
 We also presented singlet fields whose VEVs give higher order Yukawa couplings and also  can define some 4D discrete symmetries. Obtaining $\Z_{4R}$ from these singlet VEVs is desirable, which will be a future communication.

\acknowledgments{\noindent This work is supported in part  by the National Research Foundation (NRF) grants  NRF-2018R1A2A3074631.}
      



\begin{thebibliography}{99}
\def\prp#1#2#3{{Phys.\,Rep.}  {\bf #1} (#3) #2}
\def\rmp#1#2#3{{ Rev. Mod. Phys.}  {\bf #1} (#3) #2}
\def\npb#1#2#3{{ Nucl.\,Phys.\,B}   {\bf #1} (#3) #2}
\def\plb#1#2#3{{Phys.\,Lett.\,B}   {\bf #1} (#3) #2}
\def\prd#1#2#3{{Phys.\,Rev.\,D}  {\bf #1} (#3) #2}
\def\prl#1#2#3{{ Phys.\,Rev.\,Lett.}   {\bf #1} (#3) #2}
\def\err#1#2#3{ {\bf #1}   {\bf #1} (#3) #2\,(E)}
\def\jhep#1#2#3{{ JHEP}   {\bf #1} (#3) #2}
\def\jcap#1#2#3{{ JCAP}   {\bf #1} (#3) #2}
\def\zp#1#2#3{{ Z.\,Phys.}  {\bf #1} (#3) #2}
\def\epjc#1#2#3{{ Euro.\,Phys.\,J.\,C}  {\bf #1} (#3) #2}
\def\jpg#1#2#3{{J.\,Phys.\,G}   {\bf #1} (#3) #2}
\def\ijmpa#1#2#3{{ Int.\,J.\,Mod.\,Phys.\,A}   {\bf #1} (#3) #2}
\def\mpla#1#2#3{{Mod.\,Phys.\,Lett.\,A}   {\bf #1} (#3) #2}
\def\apj#1#2#3{{Astrophys.\,J.}   {\bf #1} (#3) #2}
\def\nat#1#2#3{{Nature}   {\bf #1} (#3) #2}
\def\sjnp#1#2#3{{ Sov.\,J.\,Nucl.\,Phys.}  {\bf #1} (#3) #2}
\def\apj#1#2#3{{Astrophys.\,J.}   {\bf #1} (#3) #2}
\def\mnra#1#2#3{{ Mon.\,Not.\,Roy.\,Astron.\,Soc.}   {\bf #1} (#3) #2}
\def\jetpl#1#2#3{{JETP\,Lett.}   {\bf #1} (#3) #2}
\def\pthp#1#2#3{{Prog.\,Theor.\,Phys.}  {\bf #1} (#3) #2}
\def\jkps#1#2#3{{J.\,Korean\,Phys.\,Soc.}  {\bf #1} (#3) #2}
\def\dum#1#2#3{  {\bf #1} (#3) #2}

\def\ibid#1#2#3{{\it ibid.} {\bf #1} (#3) #2}
\def\err#1#2#3{\ {\bf #1} (#3s
s
) #2\,(E)}
\def\err#1#2#3{\ {\bf #1} (#3) #2\,(E)}
  
 \bibitem{GQW74} H. Georgi, H.\,R. Quinn, and S. Weinberg,
 {Hierarchy of interactions in unified gauge theories},
   \prl{33}{451}{1974} [doi:  10.1103/PhysRevLett.33.451].
  
 \bibitem{PS73} J.\,C. Pati and Abdus Salam,
  {Unified lepton-hadron symmetry and a gauge theory of the basic interactions},
  \prd{8}{1240}{1973} [doi:  10.1103/PhysRevD.8.1240].
 
 \bibitem{GG74} H.\,M. Georgi and S.\,L. Glashow,
  {Unity of all elementary particle forces},
  \prl{32}{438}{1974} [doi:  10.1103/PhysRevLett.32.438].
  
\bibitem{Bouchiat72} C. Bouchiat, J. Ilipoulos, and P. Meyer,
{An Anomaly Free Version of Weinberg's Model},
\plb{38}{519}{1972} [doi: 10.1016/0370-2693(72)90532-1].

\bibitem{Jackiw72} D. J. Gross and R. Jackiw,
 {Effect of anomalies on quasirenormalizable theories},
\prd{6}{477}{1972} [doi:10.1103/PhysRevD.6.477].
   
\bibitem{Georgi79} H. Georgi,
 {Towards a grand unified theory of flavor},
   \npb{156}{126}{1979} [doi:  10.1016/0550-3213(79)90497-8].

\bibitem{Kim80PRL} J. E. Kim,
 {A Model of Flavor Unity},
\prl{45}{1916}{1980} [doi:10.1016/0370-2693(80)90140-9].

\bibitem{Frampton80} P. H. Frampton,
 {Unification of Flavor}, 
\plb{89}{352}{1980} [doi:10.1016/0370-2693(80)90140-9].

\bibitem{FramptonKim20} P. H. Frampton, S. K. Kang, J. E. Kim, and S. Nam,
 { $L_{\mu}-L_{\tau}$ effects to quarks and leptons from 
 flavor unification}, \prd{102}{013005}{2020}
[e-print:2004.08234 [hep-ph]].

\bibitem{GHMR1} D. J. Gross, J. A. Harvey,  E. J. Martinec, and R. Rohm, {The heterotic string},
  \prl{54}{502}{1985} [doi: 10.1103/PhysRevLett.54.502].


\bibitem{IKNQ} L. E. Ibanez, J. E. Kim, H. P. Nilles, and F. Quevedo, {Orbifold compactifications with three families of SU(3)$\times$SU(2)$\times$U(1)$^n$},
 \plb{191}{292}{1987} [doi:10.1016/0370-2693(87)90255-3].

\bibitem{Munoz88} C. Casas and C. Munoz,
 {Three generation SU(3)$\times$SU(2)$\times$U(1)$_Y$ models from orbifolds},
 \plb{214}{63}{1988} [doi:10.1016/0370-2693(88)90452-2].

\bibitem{Lykken96}  S. Chaudhuri, G. Hockney, and J. D. Lykken, 
  {Three generations in the fermionic construction}, 
  \npb{469}{357}{1996} [arXiv:hep-th/9510241].

\bibitem{PokorskiW99}  W. Pokorski and G. G. Ross,
  {Flat directions, string compactification and three generation models},
  \npb{551}{515}{1999} [e-print:hep-ph/9809537].
 
\bibitem{Cleaver99} G. B. Cleaver, A. E. Faraggi, and D. V. Nanopoulos,
  {String derived MSSM and M theory unification},
   \plb{455}{135}{1999} [e-print:hep-ph/9811427].

\bibitem{Cleaver01} G. B. Cleaver, A. E. Faraggi, and D. V. Nanopoulos,
  {A minimal superstring standard model I: Flat directions},
 IJMPA {\bf 16}, 425 (2001) [e-print:hep-ph/9904301].

\bibitem{CleaverNPB} G. B. Cleaver, A. E. Faraggi, D. V. Nanopoulos, and J. W. Walker,
  {Phenomenological study of a minimal superstring standard model},
  \npb{593}{471}{2001} [e-print:hep-ph/9910230].

\bibitem{Raby05} T. Kobayashi, S. Raby, R-J. Zhang,  {Searching for realistic 4d string models with a Pati-Salam symmetry: Orbifold grand unified theories from heterotic string compactification on a Z6 orbifold}, \npb{704}{3}{2005},  [e-print:hep-ph/0409098];   {Searching for realistic 4d string models
with a Pati–Salam symmetry. Orbifold grand unified theories from heterotic string compactification on a Z6 orbifold}, \npb{704}{3}{2005} [e-print:hep-ph/0409098].
 
\bibitem{Donagi02} R. Donagi, B. A. Ovrut, T. Pantev, and D. Waldram,
 {Spectral involutions on rational elliptic surfaces},
   Adv. Theor. Math. Phys. {\bf 5} (2002) 93 [e-print:math/0008011].
 
\bibitem{Donagi05} R. Donagi, Y-H. He, B. A. Ovrut,  and T. Pantev,
 {A heterotic standard model},
  \plb{618}{252}{2005}   [e-print:hep-th/0501070].

\bibitem{He05} R. Donagi, Y-H. He, B. A. Ovrut,  and R. Reinbacher,
  {The spectra of heterotic standard model vacua},
    \jhep{06}{070}{2005}   [e-print:hep-th/0411156].

\bibitem{Donagi06}  V. Bouchard and R. Donagi,
 {An SU(5) heterotic standard model},
  \plb{633}{783}{2006}   [e-print:hep-th/0512149].

\bibitem{He06}  V. Braun,  Y-H. He, B. A. Ovrut,  and T. Pantev,
 {The exact MSSM spectrum from string theory},
  \jhep{05}{043}{2006}   [e-print:hep-th/0512177].

\bibitem{Blumenhagen06}  R. Blumenhagen, S. Moster, and T. Weigand,
  {Heterotic GUT and standard model vacua from simply connected Calabi-Yau manifolds},
  \npb{751}{186}{2006} [e-print: hep-th/0603015].

\bibitem{Cvetic06} V. Bouchard, M. Cvetic, and R. Donagi,
 {Tri-linear couplings in an heterotic minimal supersymmetric standard model},
  \npb{745}{62}{2006} [e-print: hep-th/0602096].  

\bibitem{Blumenhagen07} R. Blumenhagen, S. Moster, R. Reinbacher, and T. Weigand,
  {Massless spectra of three generation U(N) heterotic string vacua},
  \jhep{0705}{041}{2007} [e-print:hep-th/0612039].  

\bibitem{KimJH07} J. E. Kim, J-H. Kim, and B. Kyae,
 {Superstring standard model from Z(12-I) orbifold compactification with and without exotics, and effective R-parity},
 \jhep{0706}{034}{2007} [e-print:hep-ph/0702278].

\bibitem{Faraggi07} A. E. Faraggi, C. Kounas, and J. Rizos,
  {Chiral family classification of fermionic Z2 x Z2 heterotic orbifold 
  models},
  \plb{648}{84}{2007} [e-print: hep-th/0606144].

\bibitem{Cleaver07} G. B. Cleaver,
  {In search of the (minimal supersymmetric) standard model string},
  e-print:hep-ph/0703027.

\bibitem{Munoz07} C. Munoz,
  {A kind of prediction from string phenomenology: Extra matter at low energy},
  \mpla{22}{989}{2007}  [e-print:0704.0987 [hep-ph]].

\bibitem{Lebedev08}
  O. Lebedev, H.\,P. Nilles, S. Raby, S. Ramos-Sanchez, M. Ratz,
P.\,K.\,S. Vaudrevange, and A. Wingerter,
 {The heterotic road to the MSSM with R parity},
  \prd{77}{046013}{2008} 
  [e-print:0708.2691 [hep-th]].
  
 \bibitem{Nibbelink07} S. G. Nibbelink, O.  Loukas, and F.  Ruehle, 
(MS)SM-like models on smooth Calabi-Yau manifolds from all threeheterotic string theories,
Progress of Phys. {\bf 63} (2015) 609--632
 [e-print:1507.07559v3 [hep-th]].
   
  \bibitem{Nilles08}
O. Lebedev, H.\,P. Nilles,  S. Ramos-Sanchez, M. Ratz, and
P.\,K.\,S. Vaudrevange,
  {Heterotic mini-landscape  (II): Completing the search for MSSM vacua in a Z(6) orbifold},
   \plb{668}{331}{2008} [e-print:0807.4384 [hep-th]].
   
 \bibitem{Vaudrevange11} H. P.  Nilles, S. Ramos-Sanchez, P.\,K\,S. Vaudrevange, A. Wingerter, 
 The Orbifolder: A Tool to study the Low Energy Effective Theory of Heterotic Orbifolds,
 Comp. Phys. Comm. {\bf 183} (2012) 1363--1380 [e-print:1110.5229[hep-th]].
   
\bibitem{Ginsparg87} P. H. Ginsparg,
  {Gauge and Gravitational Couplings in Four-Dimensional String 
  Theories},
\plb{197}{139}{1987} [doi:10.1016/0370-2693(87)90357-1].
  
\bibitem{Ellis89} I.  Antoniadis, J.\,R. Ellis, J\,.S. Hagelin, and D.\,V. Nanopoulos,
  {The flipped SU(5) x U(1) string model revamped},
  \plb{231}{65}{1989} [doi: 10.1016/0370-2693(89)90115-9].
 
 \bibitem{KimKyae07} J.\,E. Kim and B. Kyae,
  {Flipped SU(5) from Z(12-I) orbifold with Wilson line},
   \npb{770}{47}{2007} [e-print: hep-th/0608086].
   
 \bibitem{Weinberg79} S. Weinberg, \emph{Baryon and lepton nonconserving processes}, \prl{43}{1566}{1979} [doi:10.1103/PhysRevLett.43. 1566]. 

\bibitem{DKN84} J.\,P. Derendinger, J.\,E. Kim, and D.\,V. Nanopoulos,
  {Anti-SU(5)},
  \plb{139}{170}{1984} [doi:10.1016/0370-2693(84)91238-3 ].

\bibitem{Barr82} S.\,M. Barr,
  {A new symmetry breaking pattern for SO(10) and proton decay},
 \plb{112}{219}{1982} [doi: 10.1016/0370-2693(82)90966-2].
 
\bibitem{Veneziano82} Y. Meurice and G. Veneziano,
 SUSY vacua versus chiral fermions,
  \plb{141}{69}{1984} [doi : 10 .1016 /0370 -2693(84 )90561 -6].
  
\bibitem{Huh09} J.\,H. Huh, J.\,E. Kim, and B. Kyae,
  {SU(5)$_{\rm flip}$ x SU(5)$'$ from Z(12-I)},
   \prd{80}{115012}{2009} [e-print: 0904.1108 [hep-ph]].
 
\bibitem{KimKyae19}
 J. E. Kim and B. Kyae,
 {A model of dynamical SUSY breaking},
\plb{797}{134807}{2019} [e-print:1904.07371 [hep-th]].

\bibitem{Kimjhep15} J. E. Kim,
 {Towards unity of families: anti-SU(7) from Z(12-I)
  orbifold compactification},
\jhep{1516}{114}{2015} [e-print:1503.03104 [hep-ph]].

\bibitem{Kim19Rp} J. E. Kim,
 {R-parity from string compactification},
\prd{99}{093004}{2019} [e-print:1810.10796 [hep-ph]].

\bibitem{PHFKim20}
P. H. Frampton, J. E. Kim, S-J. Kim, and S. Nam,
 {Tetrahedral  $A_4$   symmetry in anti-SU(5) GUT},
\prd{101}{055022}{2020} [e-print:2001.02954 [hep-ph]].

\bibitem{KimKim20} J. E. Kim and S-J. Kim,
 {On the progenitor quark mass matrix},
\jkps{77}{10-16}{2020} [eprint:2004.04789 [hep-ph]].

\bibitem{DHVW1} L. J. Dixon, J. A. Harvey, C. Vafa, and E. Witten,
 {Strings on orbifolds},
 \npb{261}{678}{1985} [doi: 10.1016/0370-2693(87)90066-9]. 

\bibitem{DHVW2} L. J. Dixon, J. A. Harvey, C. Vafa, and E. Witten,
 {Strings on orbifolds (II)},
\npb{274}{285}{1986} [doi:/10.1016/0550-3213(86)90287-7].

\bibitem{INQ87} L.\,E. Iba\~nez, H.\,P. Nilles, and F. Quevedo,
  {Orbifolds and Wilson lines},
   \plb{187}{25}{1987} [doi: 10.1016/0370-2693(87)90066-9].
 
\bibitem{LNP696} K.-S. Choi and J. E. Kim,
 {\it Quarks and Leptons from Orbifolded Superstring},
  Lecture Notes in Physics, Vol. 696 [Springer, Berlin, 2006].
 
\bibitem{LNP954}  K.-S. Choi and J.\,E. Kim, {\it Quarks and Leptons from Orbifolded Superstring, 2nd Ed.},  Lecture Notes in Physics Vol. 954 (Springer-Verlag, 2020).  

\bibitem{SO10Georgi} H. Georgi,
  {The state of the art -- Gauge theories},
AIP Conf. Proc. 23 (1975) 575-582  [doi: 10.1063/1.2947450].

\bibitem{SO10Fritzsch}  H. Fritzsch and P. Minkowski, {Unified interactions of leptons and hadrons},
 Annals Phys. 93 (1975) 193  [doi: 10.1016/0003-4916(75)90211-0].
 
\bibitem{Dimopoulos81} S. Dimopoulos and F. Wilczek,
  {The unity of the fundamental interactions},
   Conference: C81-07-31 ed. A. Zichichi (Plenum, New York, 1983), p.817 [Proceedings, 19th Course of the International School of Subnuclear Physics, Erice, Italy, July 31 - August 11, 1981] [doi:  10.1007/978-1-4613-3655-6].

\bibitem{Missing} K.\,S. Babu and S.\,M. Barr,
  {Natural gauge hierarchy in SO(10)},
   \prd{50}{3529}{1994} [doi:  hep-ph/9402291].

\bibitem{Frampton79} P. Frampton,
  {SU(N) grand unification with several quark - lepton generations},
   \plb{88}{299}{1979} [doi: 10.1016/0370-2693(79)90472-6].

\bibitem{FramptonPRL79} P. Frampton and S. Nandi,
  {SU(9) grand unification of flavor with three generations},
   \prl{43}{1460}{1979} [doi:  10.1103/PhysRevLett.43.1460].

\bibitem{KimPRL13} J.\,E. Kim, 
 {Natural Higgs-flavor-democracy solution of the $\mu$ problem of supersymmetry and the QCD axion},
 \prl{111}{031801}{2013} [arXiv:1303.1822 [hep-ph]].

\bibitem{KimPLB13}  J.\,E. Kim,
  {Abelian discrete symmetries  Z(N) and Z(N)R  from string orbifolds }, \plb{726}{450}{2013} [arXiv:1308.0344 [hep-th]].

\bibitem{FramptonPLB79} P. Frampton and S. Nandi,
 {Estimate of flavor number from SU(5) grand unification }, \plb{85}{225}{1979} [doi:  10.1016/0370-2693(79)90584-7].

\bibitem{KimNilles84} J.\,E. Kim and H.\,P. Nilles,
  {The $\mu$ problem and the strong CP problem},
   \plb{138}{150}{1984} [doi: 10.1016/0370-2693(84)91890-2].

\bibitem{CKN92} E.\,J. Chun, J.\,E. Kim, and H.\,P. Nilles,
  {A natural solution of the $\mu$ problem with a composite axion in the hidden sector},
 \npb{370}{105}{1992} [doi:  10.1016/0550-3213(92)90346-D].

\bibitem{CM93} J.\,A. Casas and C. Mu\~noz,
  {Natural solution to the $\mu$ problem},
   \plb{306}{288}{1993} [e-print: hep-ph/9302227].

\bibitem{KKK07}J.\,E. Kim, J.-H. Kim,  and B. Kyae,  {Superstring standard model from $\Z_{12-I}$ orbifold compactification with and without exotics, and effective R-parity }, \jhep{0706}{034}{2007} [e-print: hep-ph/0702228].

\bibitem{Witten81}   E. Witten,
Dynamical breaking of supersymmetry,
\npb{188}{513}{1981} [doi: 10 .1016 /0550 -3213(81 )90006 -7].
  
\bibitem{Nilles82}   H. P. Nilles,
Dynamically broken supergravity and the hierarchy problem,
\plb{115}{193}{1982} [doi: 10 .1016 /0370 -2693(82 )90642 -6].
 
\end{thebibliography}
\end{document}